\documentclass[oneside,openany]{aa}  
\usepackage{graphicx}
\usepackage{amsmath}
\usepackage{amssymb}
\usepackage{breqn}
\usepackage{gensymb}
\usepackage[colorinlistoftodos]{todonotes}
\usepackage[colorlinks=true, allcolors=blue]{hyperref}
\usepackage{booktabs}
\usepackage{tikz}
\usepackage{rotating}
\usepackage{babel}
\usepackage{threeparttable}
\usepackage{caption}
\usepackage{subcaption}
\usepackage{adjustbox}
\usepackage{txfonts}
\usepackage{dirtytalk}
\usepackage{nicefrac, xfrac}
\usepackage{array}
\usepackage[normalem]{ulem}
\usepackage{balance}

\usepackage{xspace} 
\usepackage{soul} 

\defcitealias{W22}{W22}
\defcitealias{Y24}{Y24}
\defcitealias{G23}{G23}

\def\sgra{\object{Sgr~A*}\xspace} 

\newcommand\ipole{{\tt ipole}\xspace}\newcommand\bipole{{\tt bipole}\xspace}

\newcommand\dynesty{{\tt dynesty}\xspace}

\begin{document}

   \title{Hot-spots around Sagittarius~A*}

   \subtitle{Joint fits to astrometry and polarimetry}

\titlerunning{Hot-spots around Sagittarius~A*: astrometry and polarimetry}

\authorrunning{Yfantis et al.}

    \author{A. I. Yfantis
          \inst{1}
          \and
          M. Wielgus\inst{2}
          \and
          M. A. Mo{\'s}cibrodzka\inst{1}
    }

    \institute{Department of Astrophysics / IMAPP, Radboud University, P.O. Box 9010, 6500 GL Nijmegen, The Netherlands\label{inst1} \\
    \email{a.yfantis@astro.ru.nl}
    \and Max-Planck-Institut für Radioastronomie, Auf dem Hügel 69, 53121 Bonn, Germany\label{2}
    }

   \date{Received August 2024; accepted YY}

 
  \abstract
  {Observations of Sagittarius~A* (\sgra) in near-infrared (NIR) show irregular flaring activity. Flares coincide with astrometric rotation of brightness centroid and with looping patterns in fractional linear polarization. These signatures can be explained with a model of a bright hot-spot, transiently orbiting the black hole. } 
   {We extend the capabilities of the existing algorithms to perform parameter estimation and model comparison in the Bayesian framework using NIR observations from the GRAVITY instrument, and simultaneously fitting to the astrometric and polarimetric data.}
 {Using the numerical radiative transfer code \ipole, we defined several geometric models describing a hot-spot orbiting Sgr~A*, threaded with magnetic field, and emitting synchrotron radiation. We then explored the posterior space of our models in the Bayesian framework with a nested sampling code \dynesty.}
 {We have used 11 models to sharpen our understanding of the importance of various aspects of the orbital model, such as non-Keplerian motion, hot-spot size, and off-equatorial orbit. All considered models converge to realizations that fit the data well, but the equatorial super-Keplerian model is favored by the currently available NIR dataset.}
{ We have inferred an inclination of $\sim155$~deg, which corroborates previous estimates, a preferred period of $\sim70$~minutes, and an orbital radius of $\sim12$\, gravitational radii with the orbital velocity of $\sim1.3$ times the Keplerian value. A hot-spot of a diameter smaller than 5 gravitational radii is favored. Black hole spin is not constrained well. }

   \keywords{Black hole physics -- Galaxy: center -- 
                Polarization --  Magnetic fields --
                Methods: numerical  -- Methods: statistical
               }

   \maketitle

\section{Introduction}
\label{sec:intro}
The supermassive black hole (SMBH) in our Galactic Center, Sagittarius\,A* (Sgr\,A*), is a unique laboratory for studying dynamics near the black hole event horizon. With the Sgr\,A* mass of $4.3 \times 10^6\,M_{\odot}$ \citep{Gravity2022,SgraP1} the corresponding dynamical timescale at the innermost stable circular orbit (ISCO) is about 30\,min assuming Schwarzschild spacetime. This timescale sets our approximate expectations for the compact source variability, if it can be attributed to a physical orbiting component, rather than to a pattern motion \citep[e.g.,][]{matsumoto20, Conroy:2023kec}.

Sgr\,A* exhibits frequent high energy outbursts, during which the near-infrared (NIR) flux density grows by 1-2 orders of magnitude \citep{gravity:2020c}. It has been proposed that the observed NIR flares correspond to an energised orbiting component -- a hot-spot \citep[e.g.,][]{Brod-Loeb2006, Trippe2007, Hamaus2009, Tiede2020}. Since 2018, GRAVITY Collaboration observes astrometric motion of the Sgr~A* brightness centroid and polarimetric signatures associated with flares \citep{gravity:2018,G23} with the Very Large Telescope Interferometer (VLTI), supporting the hot-spot interpretation. Further observational support comes from the analysis of Atacama Large Millimeter/submillimeter Array (ALMA) polarimetric data, exhibiting similar observational features, associated with an X-ray flare \citep{Wielgus2022_LC,W22,SgraP2}. 
A number of recent papers attempted to explain the detailed physics behind formation of these polarimetric signatures of orbital motion \citep{Gelles:2021,Vos:2022,Vincent2023} and to use the phenomenological hot-spot models to interpret NIR \citep{gravityMichi, gravityAlejandra,Aimar2023} and millimeter \citep{W22, Y24,Levis2024} data. Despite variety of modeling setups and assumptions, all these studies concluded a coherent clockwise orbital motion at radius of roughly 10\,$M$, on a low inclination near face-on orbit ($150-160$~deg), with vertical magnetic field component manifesting itself in the observations. 

On the theoretical front the hot-spot model phenomenology has been connected to the physics of magnetically arrested disk \citep[MAD;][]{Narayan2003} accretion systems, exhibiting flux eruption events resembling flaring activity of Sgr~A* \citep{dexter20,Scepi2022}, and occasionally forming transiently orbiting flux tubes -- large bubbles of electrons heated by reconnection, threaded with predominantly vertical magnetic field \citep{porth21, Ripperda2022}. Numerical simulations show that such features may produce signatures consistent with the observations \citep{Najafi2024}.

In this paper we further develop our Bayesian algorithm \bipole \footnote{\bipole (Bayesian \ipole) refers to the combination of \ipole \citep{Monika2018} with \dynesty \citep{Speagle:2019ivv}.}, \citealt{Y24} (hereafter \citetalias{Y24}), previously used for fitting to polarimetric millimeter observations of ALMA, in order to simultaneously fit to the polarimetric and astrometric NIR data of GRAVITY presented in \citealt{G23} (hereafter \citetalias{G23}). We estimate the parameters of the hot-spot motion in order to compare them with reported results from GRAVITY \citepalias{G23} and ALMA \citep{W22}, as well as to discuss model selection in the Bayesian framework. Our main findings are that the model constrains dynamical parameters of the hot-spot orbit, additionally showing preference for super-Keplerian orbital velocity and for a small hot-spot size. Black hole spin is not constrained well and the more complicated models, involving background emission or off-equatorial orbit, are not favored.

\section{Methods}
\label{sec:methods}
In this section we describe the data, the model, and the fitting algorithm used to perform parameter estimation. The setup closely resembles the one developed in \citetalias{Y24} for the millimeter wavelength data analysis, including accounting for the finite light velocity ("slow light") and incorporating a complete model of the radiative transfer. Hence, in this paper we focus on presenting differences between the specific applications to ALMA and to GRAVITY datasets.
\begin{table*}[tbh!]
\begin{center}
     \caption{Parameters and prior ranges used for the fitting. All priors are flat, except for $B_{0,s}$, for which we used a truncated log-normal prior. }
\begin{tabular}{ccl}\toprule
 \midrule 
Parameter                    & Range   & Description       \\ 
 \midrule 
$i$                    & $[90^\circ,180^\circ]$  &Viewing angle (inclination) of the observer ($i=0^\circ$ is face-on, $i=90^\circ$ is edge-on)           \\
$a_*$             & $[0.0,0.9]$    &Dimensionless black hole spin        \\
$B_{0,s}$             & $[10,10^3]$  &Magnetic field strength $(G)$\\
$r_{s}$           & $[7,15]$     &Radius of the hot-spot orbit ($M$)          \\
$\phi_{\rm cam}$         & $[0,360^\circ]$    &Location of the camera at the first point of observation (shifts light curves left-right)            \\
$LP_{\rm frac}$                        & $[0,1]$      &Dimensionless scaling of the model fractional linear polarization       \\
PA                             & $[-180^\circ,180^\circ]$   &Position angle of the black hole spin projected on the observer's screen, measured east of north          \\
$K_{\textrm{coef}}$                             & $[0.5,1.5]$   &Dimensionless coefficient defining the orbital velocity with respect to the Keplerian case \\
$ss$                       & $[1,4]$      & The size ($1 \sigma$ radius) of the hot-spot ($M$)      \\
$I_{\rm BG}$                        & $[0,5]$      &Dimensionless scaling of the background emission w.r.t. max observed brightness of the hot-spot   \\
$\theta_s$                             & $[45^\circ,135^\circ]$   &The $\theta$ angle of the spot orbit, by default $\theta = 90^\circ$ (equatorial orbit) \\
$X_{\textrm{off}}$                             & $[-20,20]$   &Astrometric offset of the BH in the X direction  ($\mu \rm{as}$) \\
$Y_{\textrm{off}}$                             & $[-20,20]$   &Astrometric offset of the BH in the Y direction ($\mu \rm{as}$)  \\     
\bottomrule 
\toprule
   \label{tab:params}
\end{tabular}
\end{center}
\end{table*}

\subsection{Data}

\begin{figure}[h!]
    \centering
    \includegraphics[width=0.891\linewidth,trim={0.2cm 0.0cm 0 0.2cm},clip]{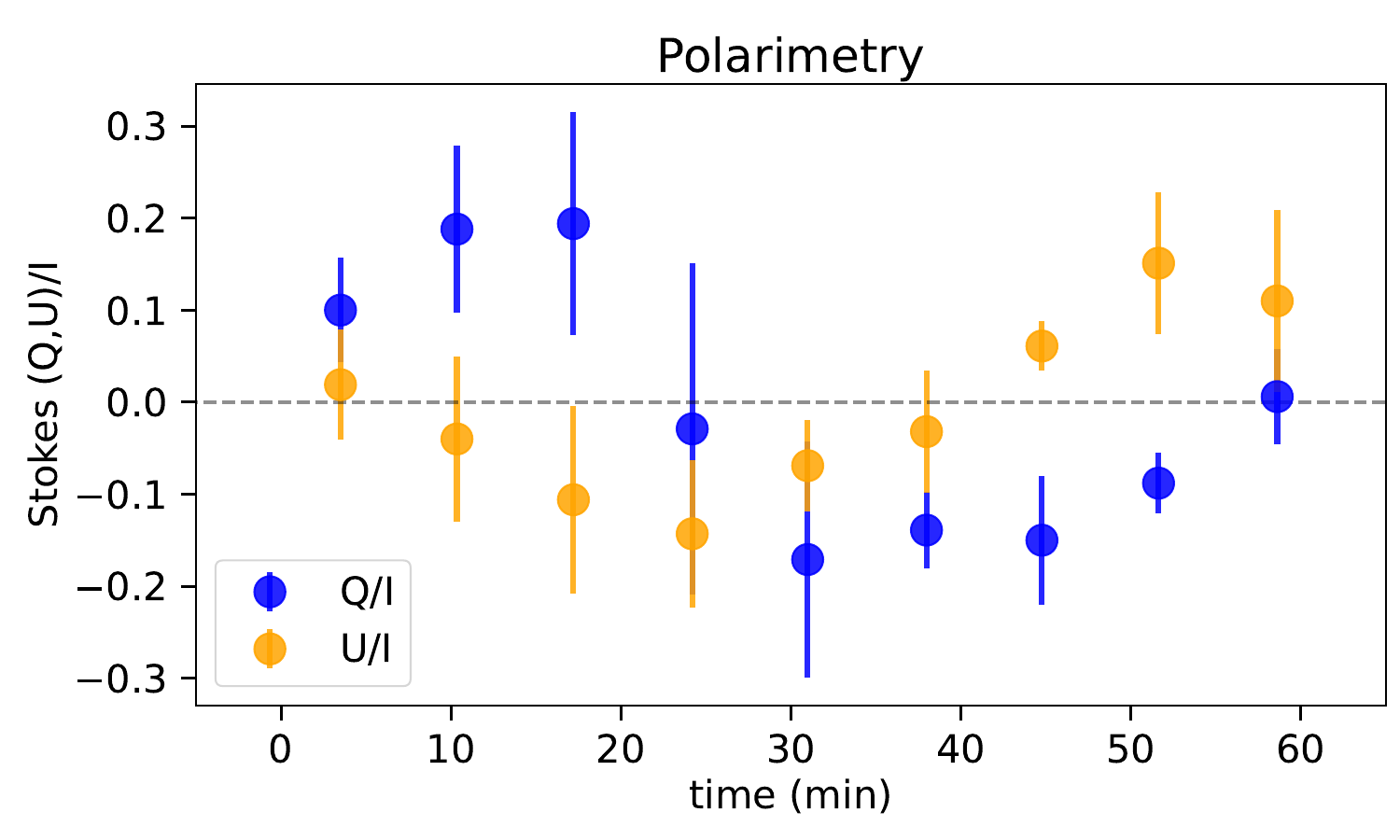} \\
    \includegraphics[width=0.891\linewidth,trim={0.2cm 0.0cm 0 0.2cm},clip]{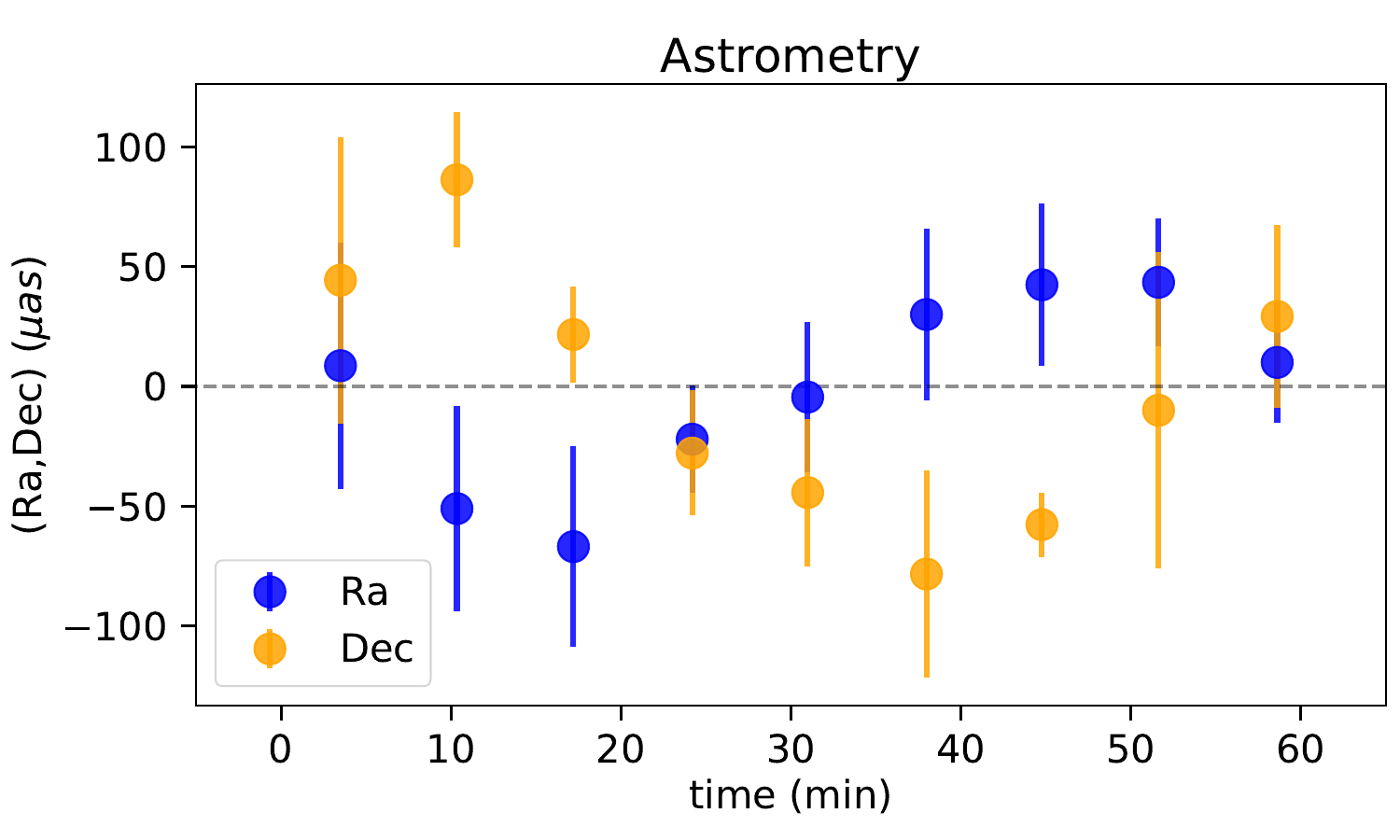} \\
    
    \caption{Ensemble averaged data set from \citetalias{G23}. Top: polarimetric data, fractional linear polarization components $\mathcal{Q}/\mathcal{I}$ and $\mathcal{U}/\mathcal{I}$ as a function of time. Bottom: astrometric data, brightness centroid position on the observer's screen $X$ (relative right ascension, Ra) and $Y$ (relative declination, Dec) as a function of time.}
    \label{fig:data}
\end{figure}

We utilized the data corresponding to the ensemble average of 8 NIR flaring events (6 with polarimetric and 4 with astrometric data available), observed by GRAVITY in 2018-2022 and first presented in \citetalias{G23}, see Fig.~\ref{fig:data}. Thus, following \citetalias{G23}, we assume that separate events can be averaged together, that is, there exist well defined mean parameters of the system, such as dynamical parameters of the hot-spot orbital motion. Furthermore, we assume that averaging between events helps to constrain these characteristic mean parameters by suppressing stochastic features of independent events.
The data differ from ALMA observations \citep{W22}, analyzed in \citetalias{Y24}, in several ways. Much shorter observing wavelength ($2.2\,\mu$m) implies negligible impact of Faraday effects and synchrotron self-absorption. These effects presented a challenge for the millimeter data analysis \citep{W22,Wielgus2023,EHT2024paper8}. Shorter wavelength and higher energy of the emitting electrons also invalidate the assumption of a constant intrinsic radiative spectral power density of the orbiter as the NIR cooling timescale is generally expected to be shorter than the dynamical timescale. Hence, in order to avoid detailed modeling the internal physics of the source \citep[e.g.,][]{Aimar2023}, we work with fractional polarization components ($\mathcal{Q}/\mathcal{I}$, $\mathcal{U}/\mathcal{I}$), robust against the changes of the intrinsic emission. What is more, the data cadence and error budgets differ significantly, as ALMA provides much higher signal to noise ratio and cadence. On the other hand, GRAVITY data include astrometric information corresponding to the relative motion of the brightness centroid, simultaneous with the polarimetric variation, while ALMA provides no astrometric data. 

\subsection{Model}

We adopted the model used in \citetalias{Y24} to enable analysis of the NIR datasets. In NIR we expect the orbiting feature to dominate over the background, given a large change of brightness between the quiescent and flaring state \citep{Do2019flare, gravity:2020c}. This is in contrast to millimeter wavelengths, for which only a~small change of brightness is present \citep[e.g.,][]{Wielgus2022_LC}. 
Hence, there is most likely no need for modelling a background disk component for the NIR case. We test this assumption with one of our hot-spot models discussed in Section~\ref{sec:results}.

On the other hand, the electron energy distribution (EED) plays a more important role in NIR than at millimeter wavelengths, and is most certainly non-thermal during the flaring state of the source. In \bipole we assume a thermal relativistic EED. While no differences are expected for the astrometric data, the EED will impact the fractional polarization \citep{Rybicki1979}. The observed fractional polarization will additionally decrease in the partially incoherent magnetic field, which seems to be the case given the values observed by GRAVITY, peaking at about 25\% \citepalias{G23}. Since this effect is difficult to be properly captured in a simple geometric model, we include an additional parameter $LP_{\rm frac}$ to scale the model fractional polarization with a constant number. Furthermore, the internal physics of the hot-spot, probed only through fractional polarization, is not constrained well and degenerates between number density, temperature (or, more generally, EED), and the magnetic field strength for the synchrotron origin of the emission. For that reason we fix the dimensionless electron temperature to a large value of $\Theta_{\rm e} = 500$, encouraging emission at the 2.2\,$\mu$m wavelength. We additionally fix the peak number density of the hot-spot to $n_{\rm e} = 10^7$ cm$^{-3}$, so the only estimated parameter related to the internal physics is the magnetic field strength $B_{0,s}$. Given the nature of the data, it can essentially be considered to be a~nuisance parameter, enforcing that the synchrotron emission extends to the considered frequency.

The hot-spot itself is modeled as a Gaussian blob with a radius ($1\, \sigma$) of $3$\,$M$ ($M$ $\equiv r_{g} = GM/c^2$, we use $G=c=1$ units when it does not lead to ambiguities) and the Gaussian tail suppressed at $1.5\, \sigma$. We assume clockwise direction of the orbital motion, which is strongly constrained by the astrometric data. The simulated hot-spot travels as a rigid body on a circular orbit in the equatorial plane, with a constant orbital velocity equal to a fraction of the Keplerian orbital speed at the hot-spot center, which is modeled with $K_{\textrm{coef}}$ parameter. It affects the period of a~circular orbit, which is given in Kerr spacetime by
\begin{equation}
    P=\frac{2\pi}{K_{\textrm{coef}}}\frac{r_g}{c}\left[\left(\frac{r}{r_g}\right)^{1.5} + a_*\right], 
\label{eq:P}
\end{equation} 
where $r = r_s$ is the distance of the spot from the black hole center, $r_g = GM/c^2$ is the gravitational radius, and $a_*=Jc/GM^2$ is the dimensionless black hole spin. These aspect of the setup follow \citet{Vos:2022} and \citetalias{Y24}, and we assume magnetic field to be vertical in the comoving frame of the hot-spot, similarly as in \citetalias{Y24} and consistent with the conclusions of \citetalias{G23}. New developments include fitting to astrometric data with the imaged hot-spot brightness centroid, fitting for the hot-spot size, joint model of a hot-spot and a constant background component, and enabling circular off-equatorial orbits. Unlike in \citetalias{G23}, our hot-spot is not a Keplerian point source, we also do not neglect the impact of higher order images\footnote{Under these simplifications the fractional polarization pattern reduces to an ideal circle on a $(\mathcal{Q}/\mathcal{I}, \mathcal{U}/\mathcal{I})$ plane.}, and we simultaneously fit to astrometric and polarimetric observations. 

\subsection{Fitting procedure}
We updated \bipole in order to enable simultaneous fitting to astrometric and polarimetric data. The software combines \ipole, which is a ray-tracing code for the covariant general relativistic transport of polarized light in curved space-times developed in \cite{Monika2018} with the nested sampling Bayesian tool \dynesty, introduced by \cite{Speagle:2019ivv} and further developed by \cite{sergey_koposov_2023_7995596}.

The resolution of ray-traced images calculated by \ipole was set to $64\times64$ pixels (see the discussion on the limitations of the resolution in \citetalias{Y24}) and the \dynesty sampling parameters were set to 800 live points with 100 extra live points for 5 batches, using the dynamical sampler, introduced by \citet{Higson:2019}. The remaining hyper-parameters were identical with the set up of \citetalias{Y24}, resulting in a single likelihood evaluation on a synthetic movie of an orbiting hot-spot calculated in $0.5$ second  using a~24 cores machine\footnote{Using the COMA computational cluster at Radboud University.}. The log-likelihood function $\mathcal{L}$ was defined in a~standard way as 
\begin{equation}
    \label{eq:LKLHD_CP}
    \mathcal{L}(\vec{p})=-\sum_j^S\frac{\left[S_j-\hat{S}_j(\vec{p})\right]^2}{2\sigma_{S,j}^2}\,,
\end{equation}
where $\vec{p}$ is the vector of the model parameters that are to be estimated, $S$ are the observed values of the 4 fitted quantities $\mathcal{Q}\,,\mathcal{U}\,,X\,,Y$, while $\hat{S}\,$ represents predictions of the model. The uncertainties $\sigma_{S,j}$ follow the values reported in \citetalias{G23}. A list of all the fitted model parameters and their prior ranges is shown in Table~\ref{tab:params}.

\begin{table*}[tbh!]
\begin{center}
     \caption{Hot-spot models used to fit Sgr~A* GRAVITY data. Left: particular configuration choices per model. Right: main findings, where the reported values correspond to the ML estimators. See Appendix \ref{app:cornplots} for the associated uncertainties.}
\begin{tabular}{llcc|ccccccc}\toprule
 \midrule 
Model ID &$B_{\rm field}$ &Keplerianity &Parameters& $\log Z$&$\chi_{\text{eff}}^2$ & $i$ (deg) & $r_s$ ($M$) &$K_{\text{coeff}}$&P (min) &PA (deg) \\ 
\midrule
K\_db   & default          & fixed &7 & -22.9&0.57& 158 & 10.0& -&72 &127 \\  
K\_fb   & flipped           & fixed  &7& -23.6&0.57 &158 &  10.0 & -&72 & 127\\
supK\_db   & default         & super  &8& -21.9&0.49 & 156 & 11.6 & 1.3&70&125\\
subK\_db   & default         & sub  &8& -24.3& 0.67& 157& 10.0 & 1.0&72&125 \\

supK\_db\_a0  & default       & super  &7 & -22.0&0.50 & 154& 13.0 & 1.4&74&127\\
subK\_db\_a0 & default       & sub  &7& -24.7&0.60 &157 & 10.3 & 1.0&72&126\\
supK\_db\_pa180   & default    & super  &7& -27.9&1.04 & 156&9.4 &1.0&60&-\\
subK\_db\_pa180  & default     & sub  &7& -28.2&1.05 & 164& 7.2& 0.5&85&-\\
supK\_db\_ss  & default     & super  &9& -21.6&0.59 & 155& 11.6&1.2&73&125 \\
supK\_db\_BG  & default     & super  &9& -24.8&0.60& 154 &12.3&1.4&69& 127\\
supK\_db\_ofst &default&super&11& -24.0&1.10 & 157 & 11.5 &1.1&71&112\\

\toprule
\label{tab:models}
\end{tabular}
\end{center}
\end{table*}

\section{Results}
\label{sec:results}

\begin{figure*}[tbh!]
    \centering
    \includegraphics[width=0.45\linewidth,trim={0.1cm 0.0cm 0 0.1cm},clip]{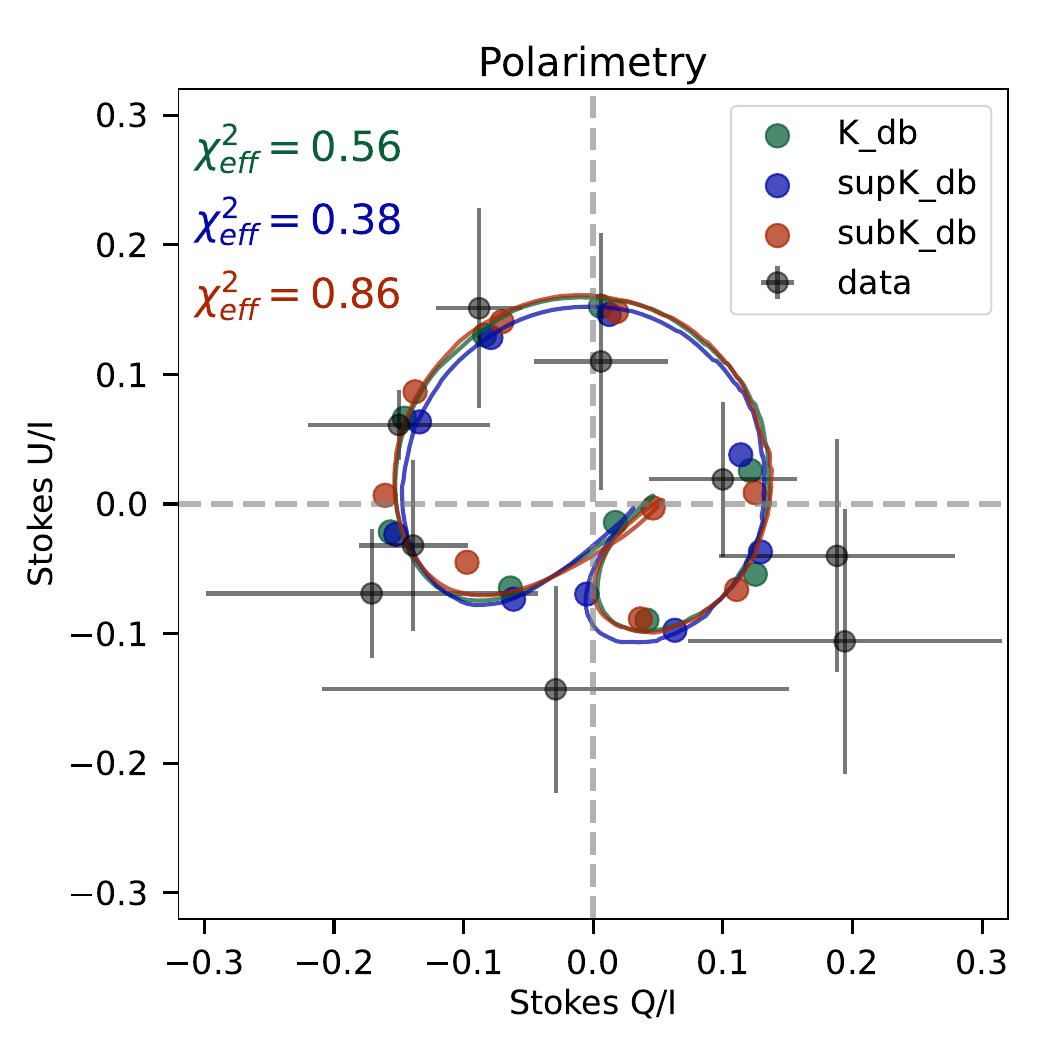} 
\includegraphics[width=0.46\linewidth,trim={0.1cm 0.0cm 0 0.1cm},clip]{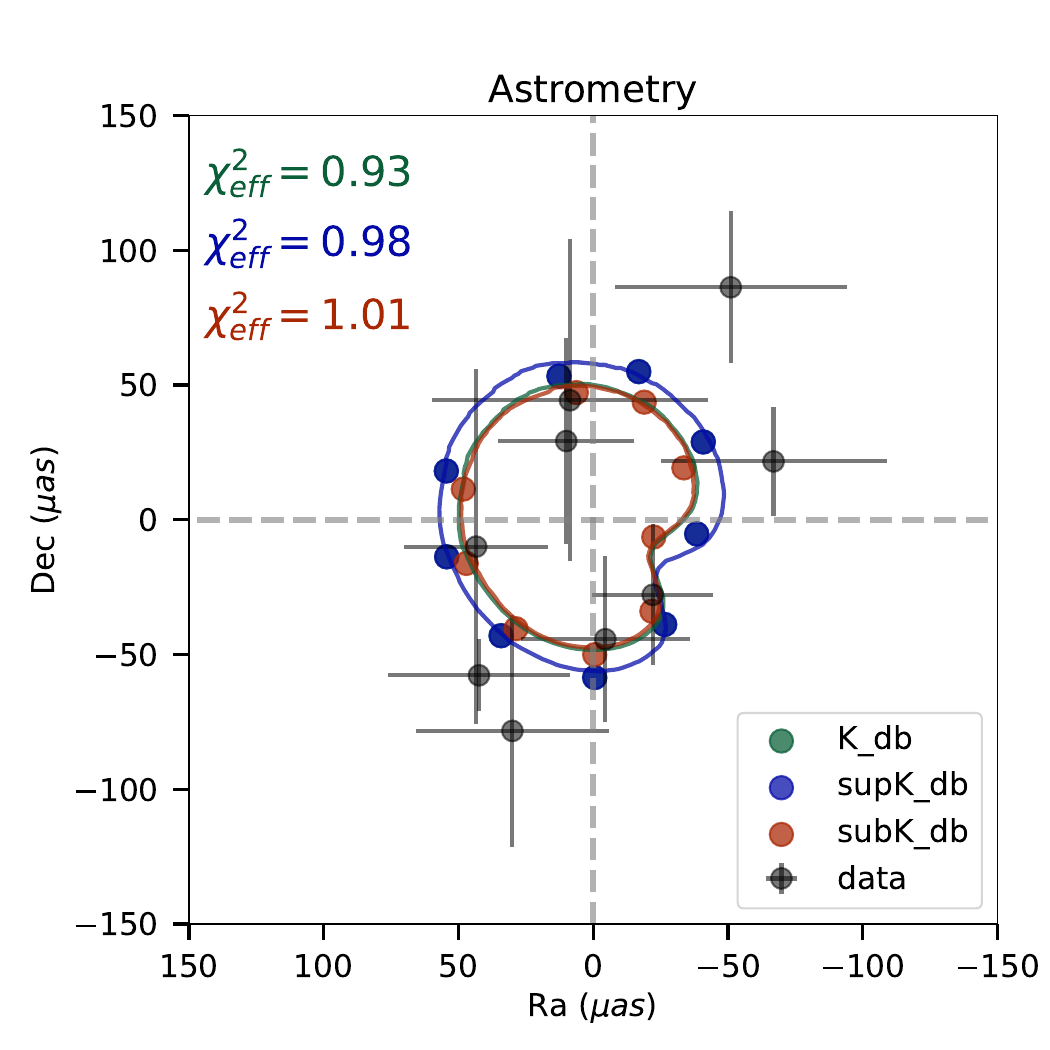} 
    \caption{Light curves from the ML estimators for models K\_db, supK\_db and subK\_db. The solid lines correspond to max-likelihood (ML) estimator, recalculated at a higher image resolution (192 $\times$ 192). $\chi^2_{\rm eff}$ is calculated separately for polarimetry and astrometry. }
    
    \label{fig:lc_nK}
\end{figure*}

We fitted 11 variants of the orbiting hot-spot model to the dataset of \citetalias{G23} in order to understand the constraints posed by the NIR observations. The models and their performance are summarized in Table~\ref{tab:models}. We used the logarithm of the Bayesian evidence (marginal model likelihood) $\log (Z)$ to compare between the models. At the level of uncertainties reported by \citetalias{G23}, all considered variants produce good quality fits to the data, as indicated by reduced effective (normalized by number of data points minus number of model parameters) $\chi_{\rm eff}^2$ metric for the maximum likelihood estimators (MLE), reported in Table~\ref{tab:models}. Our most successful models are the equatorial super-Keplerian ones without the background emission. The orbital inclination is well constrained across the models to be around $155^\circ \pm 5^\circ$, which is also consistent with constraints from \citetalias{G23} obtained with the Keplerian point source model of \citet{Gelles:2021} as well as with ALMA constraints from \citet{W22} and \citetalias{Y24}. Below we provide description of the models and discuss constraints on other parameters. Quantitative information about the results of the fitting and associated uncertainties is given in Appendix \ref{app:cornplots}.

\textit{Evidence uncertainties:} For the first test we performed the fit of an orbiting hot-spot on a~Keplerian (circular timelike geodesic) orbit, with two different polarities of the vertical magnetic field $B_{\rm field}$, models K\_db  and K\_fb. The magnetic field polarity impacts the results only through Faraday effects, which should be negligible for NIR frequencies. As expected, two models find a very similar MLE parameters. However, the estimated log-evidence differs by 0.7. We use this result to calibrate the uncertainties of the evidence, concluding that in order to claim preference of one model over another, we require log-evidence difference of at least $\Delta \log Z = 2.0$.
The models that we characterize as disfavored have $\Delta \log Z > 3.5$ with respect to best performing models. 
This corresponds to Bayes factor of $K>33$, indicating very strong preference in the model selection \citep{jeffreys1998theory}.

\textit{Orbital velocity:} We then compared the performance of super-Keplerian (supK\_db) and sub-Keplerian (subK\_db) models. There is some preference for super-Keplerian motion following the log-evidence argument (see Table~\ref{tab:models}), however, Keplerian, super-Keplerian, and sub-Keplerian MLEs all match the data well, as indicated in Fig.~\ref{fig:lc_nK}. In case of the sub-Keplerian model the ML estimator approaches Keplerian limit. There are theoretical arguments for super-Keplerian, off-equatorial orbits, e.g., \cite{matsumoto20,Lin2023,Aimar2023}. Preference for super-Keplerianity in case of equatorial orbits models was also concluded by \citet{Antonopoulou2024}. 

The quantity constrained well by the data set is the period of the motion. The astrometric uncertainties allow to trade between orbital velocity (scaled with $K_{\rm coeff}$) and orbital radius $r_s$, while keeping the period to about 70 minutes. We make a conservative estimate of the orbital radius $r_s = 11 \pm 2\, M$, consistent between the models. This value is also consistent with previous estimates, although slightly larger than the typically reported in NIR $r_s \approx 9 \,M$ \citep[][\citetalias{G23}]{gravityMichi}.

\textit{Black hole spin:} With the next test our aim was to see if the data constrain deviation of the black hole spin from a default assumption of Schwarzschild spacetime ($a_* =0$). For that purpose we test super- and sub-Keplerian models with spin fixed to zero, supK\_db\_a0 and subK\_db\_a0, respectively. Fixing the spin results in only a very small decrease of the log-evidence, less than the assumed threshold for the model selection and less than the evidence estimation uncertainties. We conclude that the black hole spin cannot be robustly inferred from the current data set and its impact on the models is much less significant than that of the non-Keplerian orbital velocity. This is further confirmed by very broad and non-constraining BH spin posteriors in all other models as seen in the corner plots presented in Appendix~\ref{app:cornplots}.

\textit{Spin axis position angle:} The parameter PA, corresponding to the position angle of the black hole spin on the observer's screen appears to be well-constrained, much like the orbital inclination (see Table~\ref{tab:models}). However, the value we find, PA $\sim 130^\circ \pm 20^\circ$, differs significantly from 180$^\circ$ reported in \citetalias{G23}. On the other hand for the ALMA dataset \citet{W22} reported 57$^\circ$ (or 237$^\circ$, given the model degeneracy) under the external Faraday screen model, which corresponds to 25$^\circ$ (or 205$^\circ$) with the internal Faraday screen assumption and \citetalias{Y24} found reasonably consistent values under the internal Faraday screen assumption \citep[see also][]{Wielgus2023}. Thus, we test models with PA fixed to 180$^\circ$ (supK\_db\_pa180 and subK\_db\_pa180). These models perform worse, and they can be disfavored based on the log-evidence criterion. We conclude that the PA can be estimated based on this data set, and that the value of 180$^\circ$ is disfavored. 

\textit{Hot-spot size:} Our next test involved varying the size of the hot-spot in the supK\_db\_ss variant of the model; this is done with the parameter $ss$ that adjusts the $1\,\sigma$ radius of the spot. By default we used a large hot-spot ($ss=3M$) to allow for lower resolution of reconstructed images and to speed up the computation time. In order to model smaller hot-spots we increase the resolution from $64 \times 64$ to $192\times 192$. We notice only marginal improvement of the log-evidence for this more general model with respect to the supK\_db model. Nonetheless, the marginalized posterior distribution indicates a preference for smaller hot-spot size. The results correspond to an upper limit on the hot-spot size at $ss = 2.16\,M$ (full width at half maximum of 5.09\,$M$) at the 68\% confidence level. This is a similar conclusion to the one given in \citet{gravityMichi}, where the hot-spot diameter was estimated to be smaller than $5\,M$.

\textit{Accretion disk component:} With the model supK\_db\_BG we investigated a presence of another emission component, associated with the background accretion disk, invoked in \citetalias{G23}, among others. We model this component as an unpolarized feature with its brightness centroid located at the origin of the coordinate system. It affects both astrometric results (by reducing the width of the observed brightness centroid loop), and the polarimetric ones (by increasing the total flux density in the denominator of the fractional polarization). Adding a background component to a~model results in a decrease of the log-evidence by 2.9, which we interpret as presence of a background emitter being discouraged by the data. 

\textit{Off-equatorial orbit:} For the final test we consider a model supK\_db\_ofst, allowing for an off-equatorial hot-spot orbit, at a~fixed Boyer-Lindquist radius $r$ and latitude $\theta$, either in the background or in the foreground of the equatorial plane. The offset results in an astrometric shift of the BH position by $(X_{\rm off}, Y_{\rm off})$, which are additional fitted parameters, making this variant our most complicated one with the total of 11 degrees of freedom (Table~\ref{tab:models}). A~motivation to consider this generalization is related to the potential presence of a shift between the astrometric location of the Sgr~A* center of mass and the center of the astrometric loops \citep{gravity:2018}, as well as hints of a super-Keplerian off-equatorial hot-spot motion \citep{matsumoto20,Aimar2023}. This model does not improve either log-evidence or $\chi^2_{\rm eff}$. Thus, our conclusion is that the available data do not support more general off-equatorial models.

\section{Summary and conclusions}
\label{sec:conclusions}
Very simple phenomenological models of an orbiting hot-spot explain well the astrometric and polarimetric NIR observations of flaring Sgr~A* obtained by VLTI / GRAVITY. Given the associated uncertainties, one overfits the data already with the basic Keplerian equatorial model. Nonetheless, the relative likelihood of the models can be studied, and constraints on different model parameters can be compared. We find that these models consistently constrain parameters of orbital radius, inclination, and the observed position angle of the axis normal to the orbital plane (spin axis). The depolarization of the orbiting component is also constrained, and the preference for a smaller size of the emitting region is identified. We also show how black hole spin is not constrained by the available dataset. Furthermore, super-Keplerian motion is favored over sub-Keplerian. We show that data are consistent with one zone emission model (only the orbiting hotspot, no background emission) and with the near equatorial orbit model ($\pm20^\circ$). More complicated models, such as the separate background component, are not justified by the constraining power of the currently available data sets.

There are excellent prospects for future observations of flaring Sgr~A* to test, validate and sharpen the orbital model. In NIR the upgraded GRAVITY instrument \citep{GravityPlus} will provide improved observing sensitivity. It is particularly interesting to study NIR flares with simultaneous (sub-)millimeter monitoring, which can be done with ALMA with full polarization, at extremely high signal to noise ratio and time cadence. Simultaneous observations with very long baseline interferometric arrays could in the near future offer to resolve the structure of the flaring source in both time and space \citep{Emami2023,Johnson2023}. Such multiwavelength observations would not only yield a strong test of the hot-spot model, they would also enable detailed studies of thermodynamics of the orbiting component to better understand the mechanism and energetics behind the phenomenon of flares. Development of dedicated data analysis pipelines, as the one presented in this paper, will be crucial to extract science from these more constraining future observations.

\begin{acknowledgements}
We thank F. Vincent, M. Bauböck, and D. Ribeiro for useful discussions. We also thank Paris Observatory in Meudon, where part of this research was conducted, for the hospitality. This publication is a part of the project Dutch Black Hole Consortium (with project number NWA 1292.19.202) of the research programme of the National Science Agenda which is financed by the Dutch Research Council (NWO). MM and AY acknowledge support from NWO, grant no. OCENW.KLEIN.113. MM also acknowledges support by the NWO Science Athena Award. 
\end{acknowledgements}
\balance
\bibliographystyle{aa} 
\bibliography{library.bib} 

\newpage

\onecolumn
\begin{appendix}

\section{Posterior distributions for the fitted models}
\label{app:cornplots}
We characterize the posterior distributions of our 11 models fitted to data with the following corner plots. Above the one-dimensional posteriors on the diagonal we provide the estimates of the fitted parameters as a median of the distribution with a 68\% confidence interval (thin dashed lines in diagonal panels). We also present the maximum likelihood estimator (thick black lines in the corner plot) and its consistency with the data (below the corner plots). The presentation of the models follows the order from Table~\ref{tab:models}.

\begin{figure*}[h!]
    \centering
     \includegraphics[width=0.69\linewidth,trim={0.2cm 0.0cm 0 0.2cm},clip]{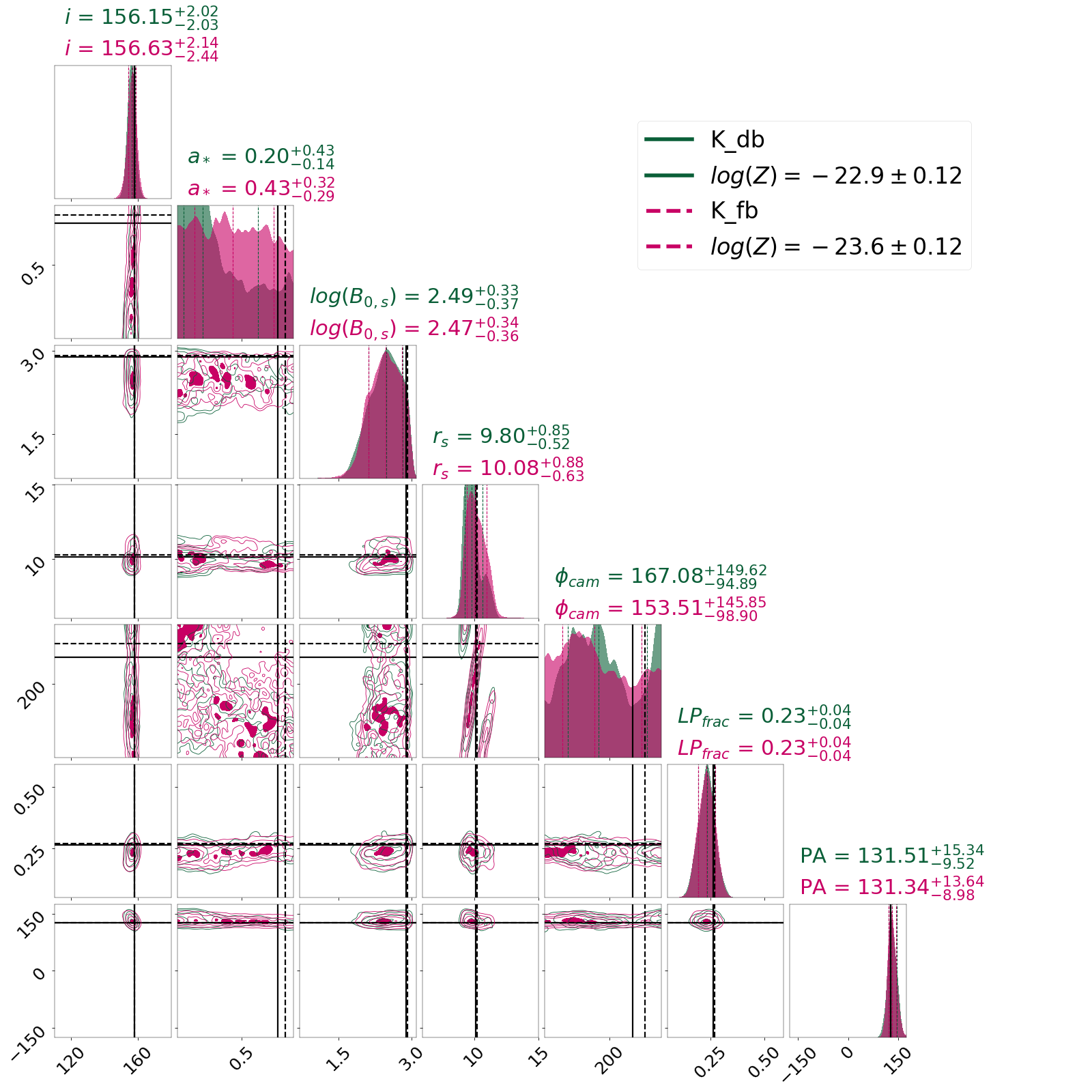} \\
    \includegraphics[width=0.392\linewidth,trim={0.1cm 0.0cm 0 0.1cm},clip]{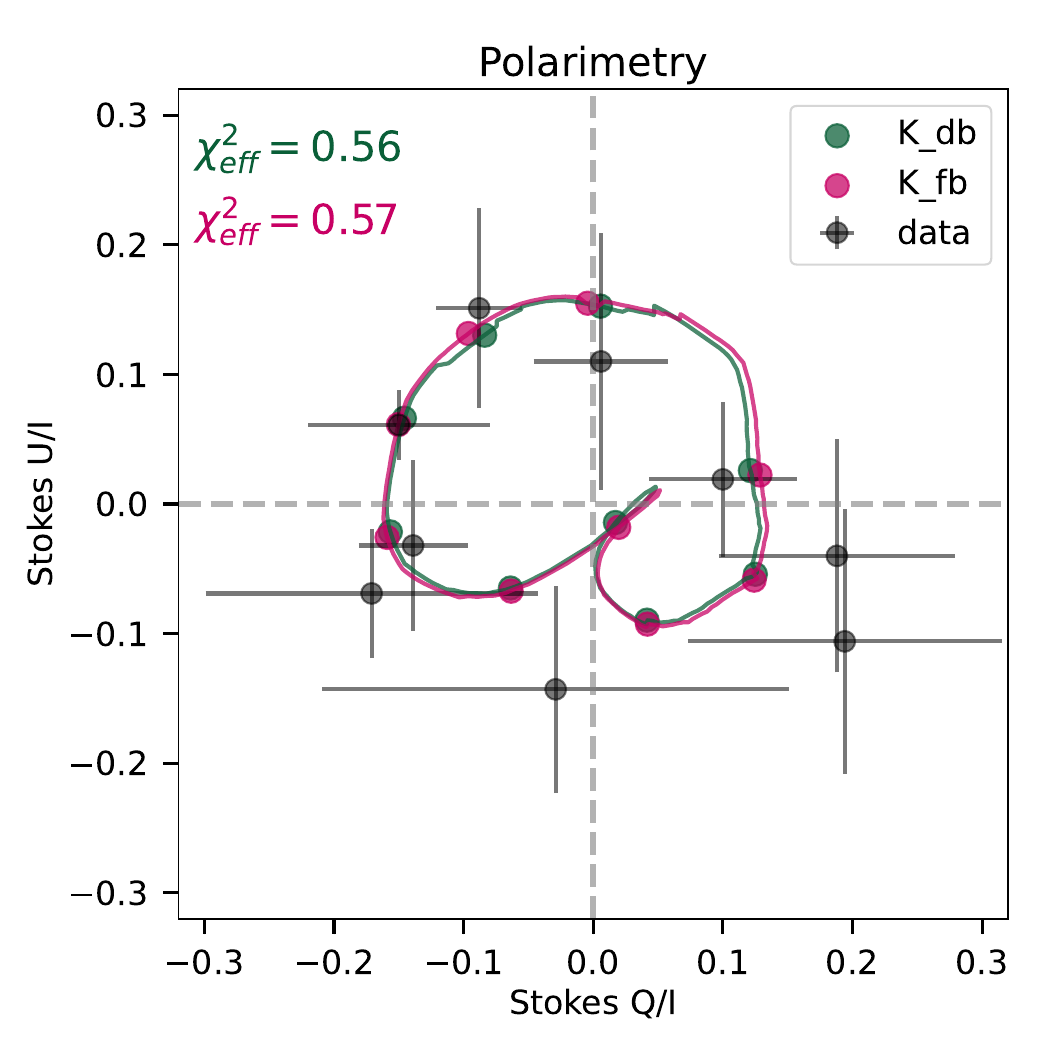} 
    \includegraphics[width=0.4\linewidth,trim={0.1cm 0.0cm 0 0.1cm},clip]{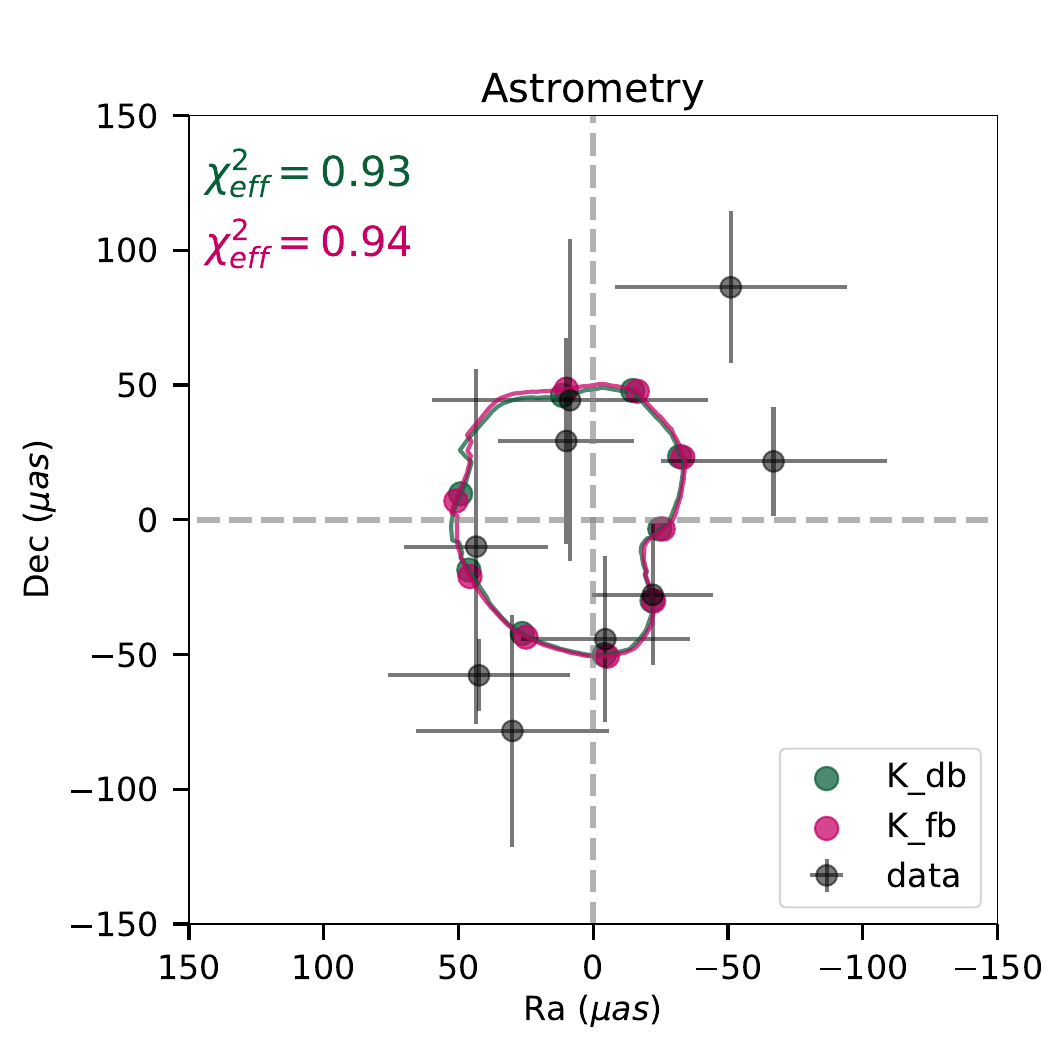} 
    \caption{Top: corner plot of the posterior distributions for all fitted parameters for the Keplerian orbital velocity models K\_db and K\_fb. The results for the two magnetic field polarities are shown in different colors, and their respected log-evidence values are reported in the legend.
    Bottom: maximum likelihood estimators compared to the data. Points denote the fitting using observational timestamps, while lines are made in post-processing for visual aid.}
    \label{fig:corner_K}
\end{figure*}

\begin{figure*}[tbh!]
    \centering
    \includegraphics[width=0.781\linewidth,trim={0.2cm 0.0cm 0 0.2cm},clip]{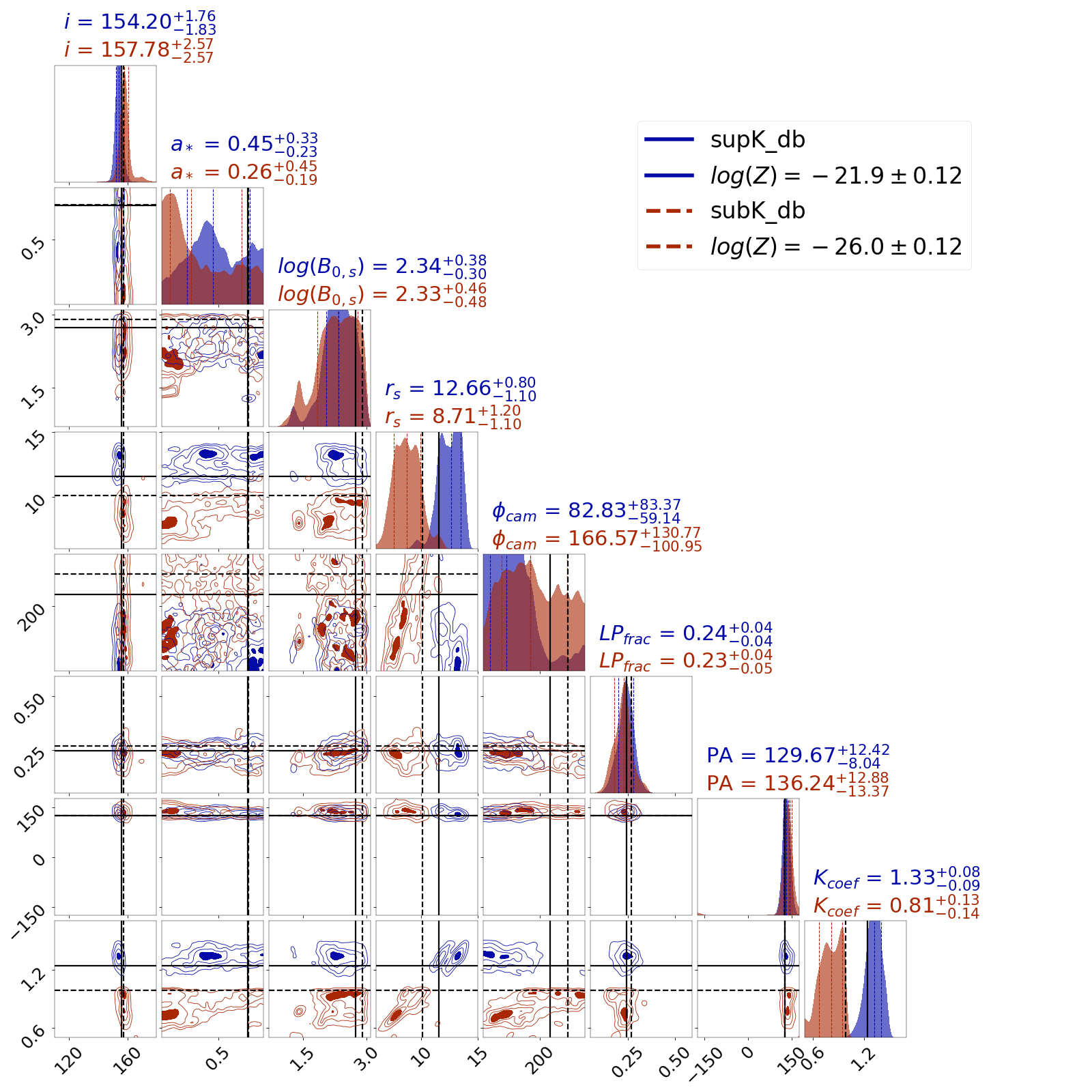} \\
    \includegraphics[width=0.45\linewidth,trim={0.1cm 0.0cm 0 0.1cm},clip]{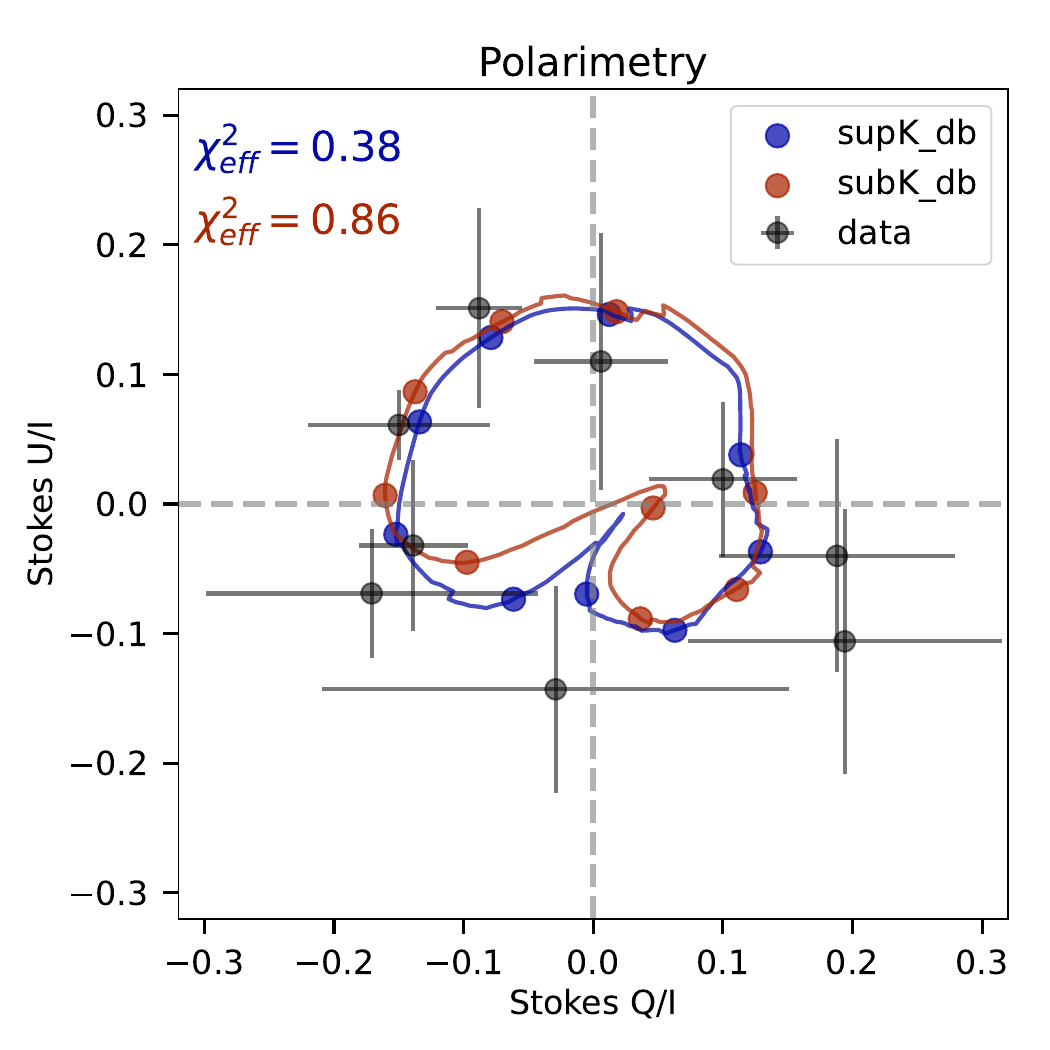} 
    \includegraphics[width=0.461\linewidth,trim={0.1cm 0.0cm 0 0.1cm},clip]{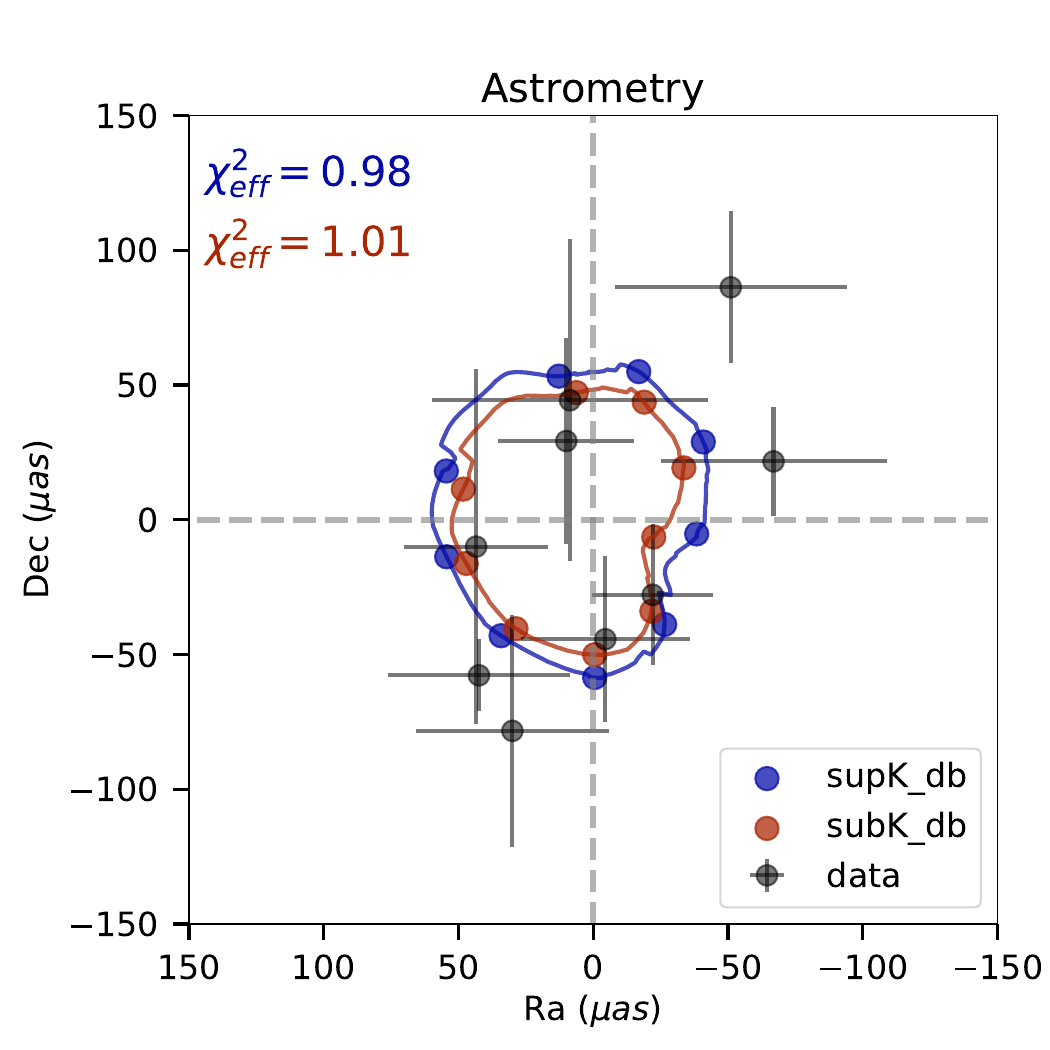} 
    \caption{Same as Fig. \ref{fig:corner_K} but for models supK\_db and subK\_db, comparing sub-Keplerian and super-Keplerian setup.}
    \label{fig:corner_nK}
\end{figure*}

\begin{figure*}[tbh!]    \centering
    \includegraphics[width=0.781\linewidth,trim={0.2cm 0.0cm 0 0.2cm},clip]{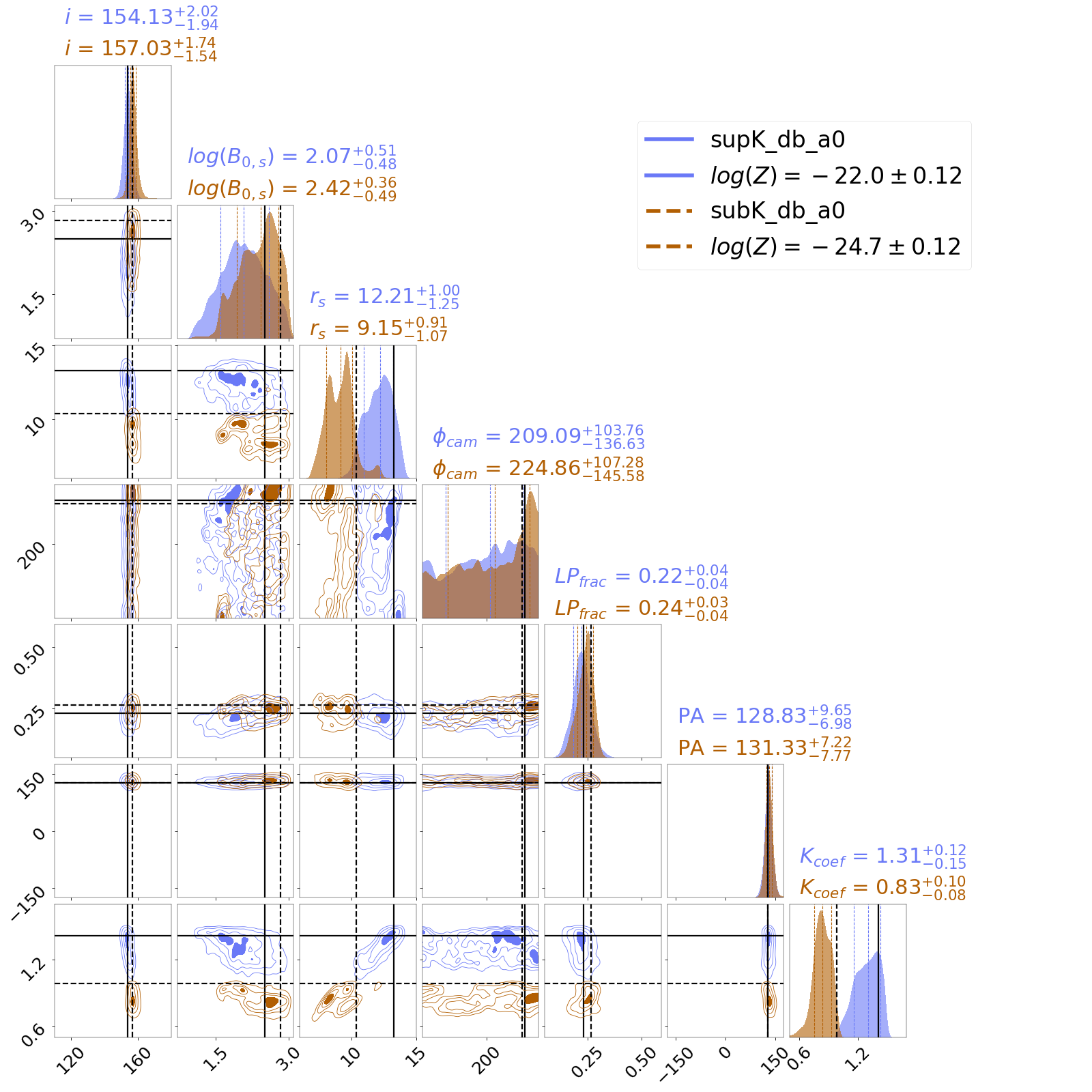} \\
    \includegraphics[width=0.45\linewidth,trim={0.1cm 0.0cm 0 0.1cm},clip]{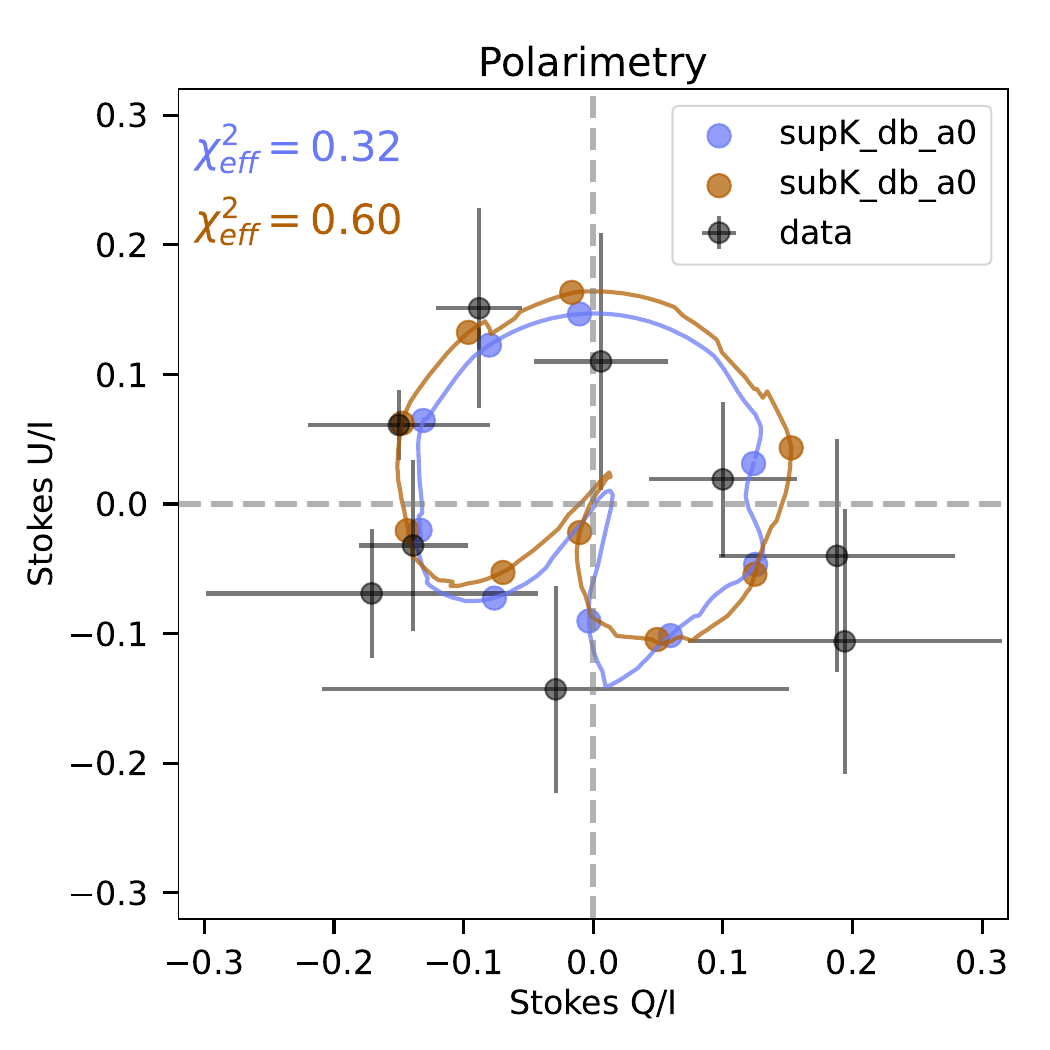} 
    \includegraphics[width=0.461\linewidth,trim={0.1cm 0.0cm 0 0.1cm},clip]{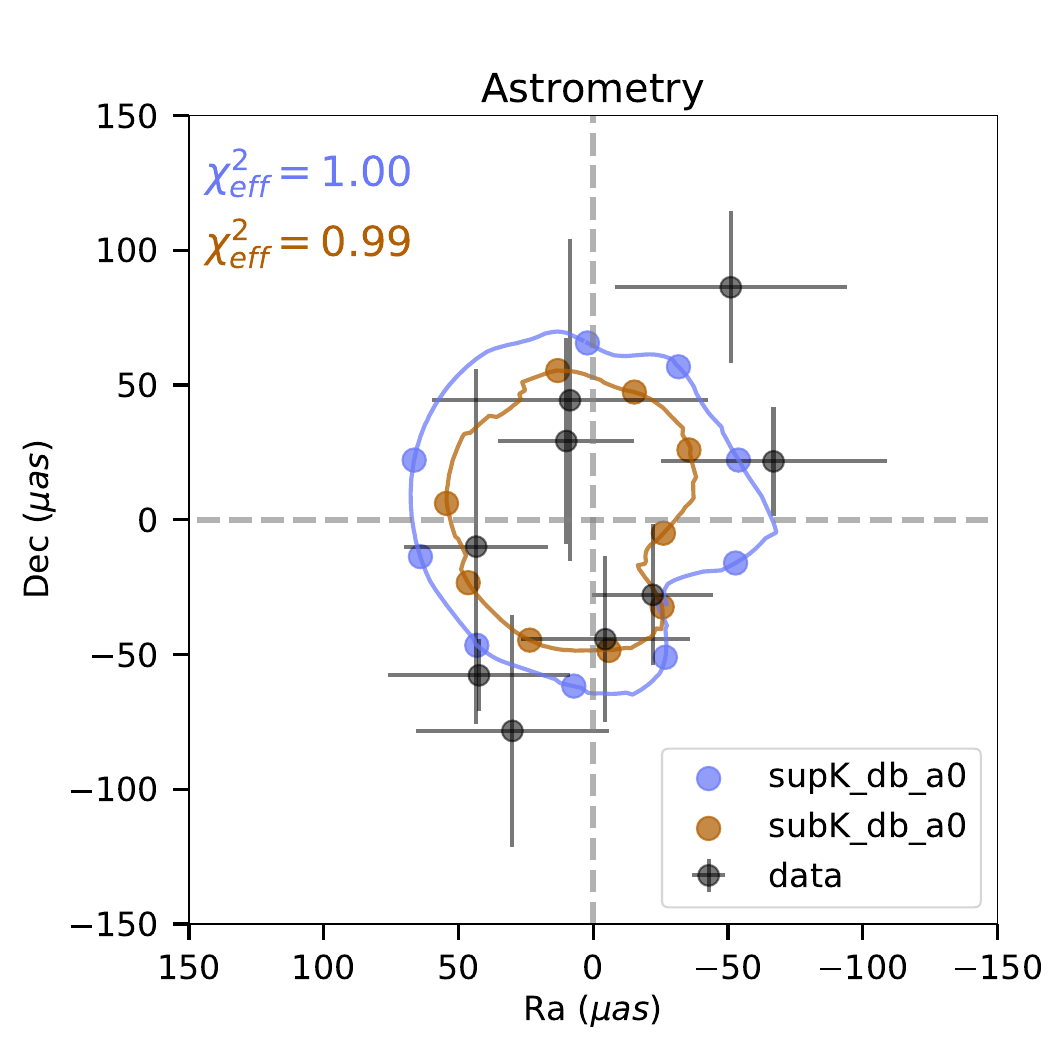} 
    \caption{Same as Fig. \ref{fig:corner_K} but for models supK\_db\_a0 and subK\_db\_a0, testing for sensitivity to the BH spin.}
    \label{fig:corner_a0}
\end{figure*}

\begin{figure*}[tbh!]
    \centering
    \includegraphics[width=0.781\linewidth,trim={0.2cm 0.0cm 0 0.2cm},clip]{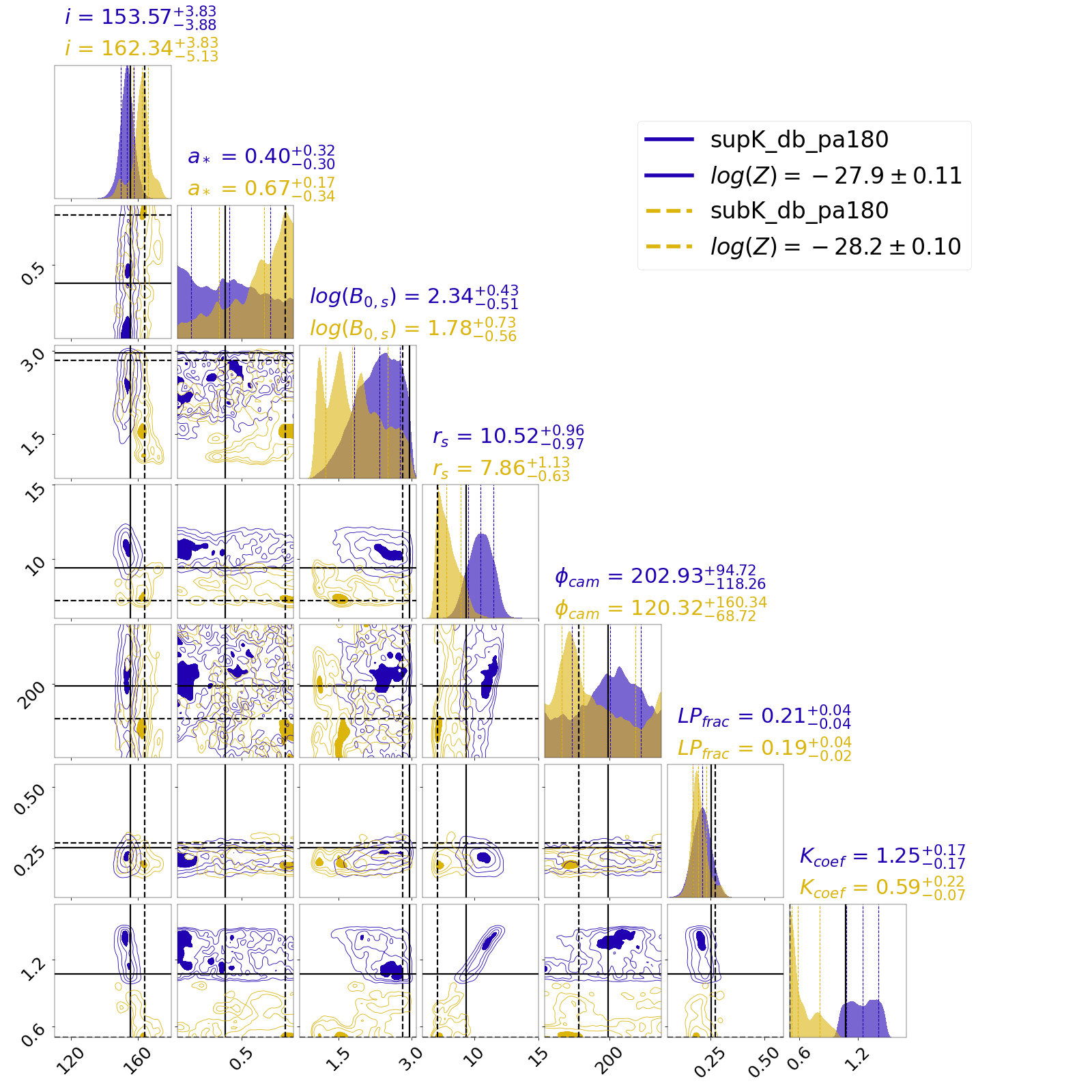} \\
    \includegraphics[width=0.45\linewidth,trim={0.1cm 0.0cm 0 0.1cm},clip]{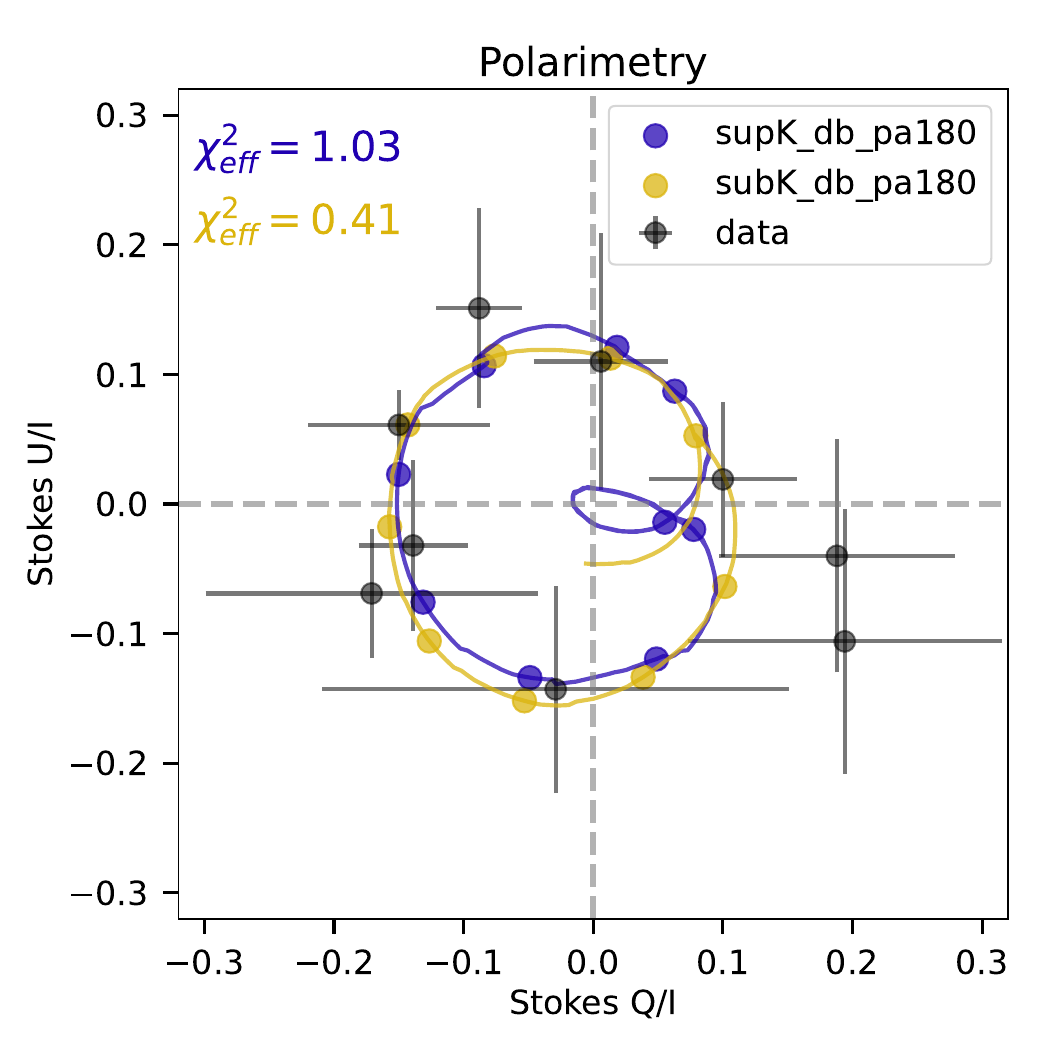} 
    \includegraphics[width=0.461\linewidth,trim={0.1cm 0.0cm 0 0.1cm},clip]{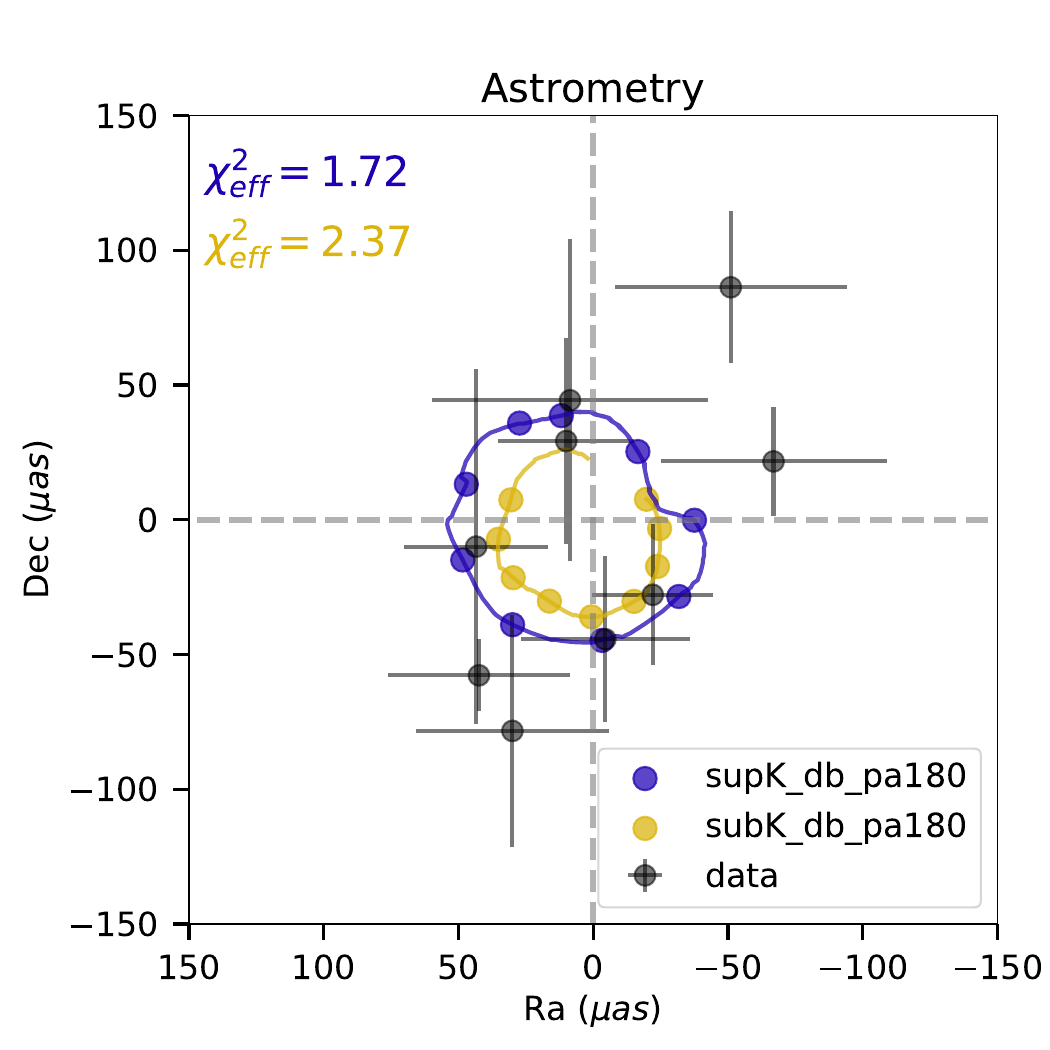} 
    \caption{Same as Fig. \ref{fig:corner_K} but for models supK\_db\_pa180 and subK\_db\_180, testing for sensitivity to the observed position angle of the spin axis.}
    \label{fig:corner_pa180}
\end{figure*}

\begin{figure*}[h!]
    \centering
    \includegraphics[width=0.76\linewidth,trim={0.2cm 0.0cm 0 0.2cm},clip]{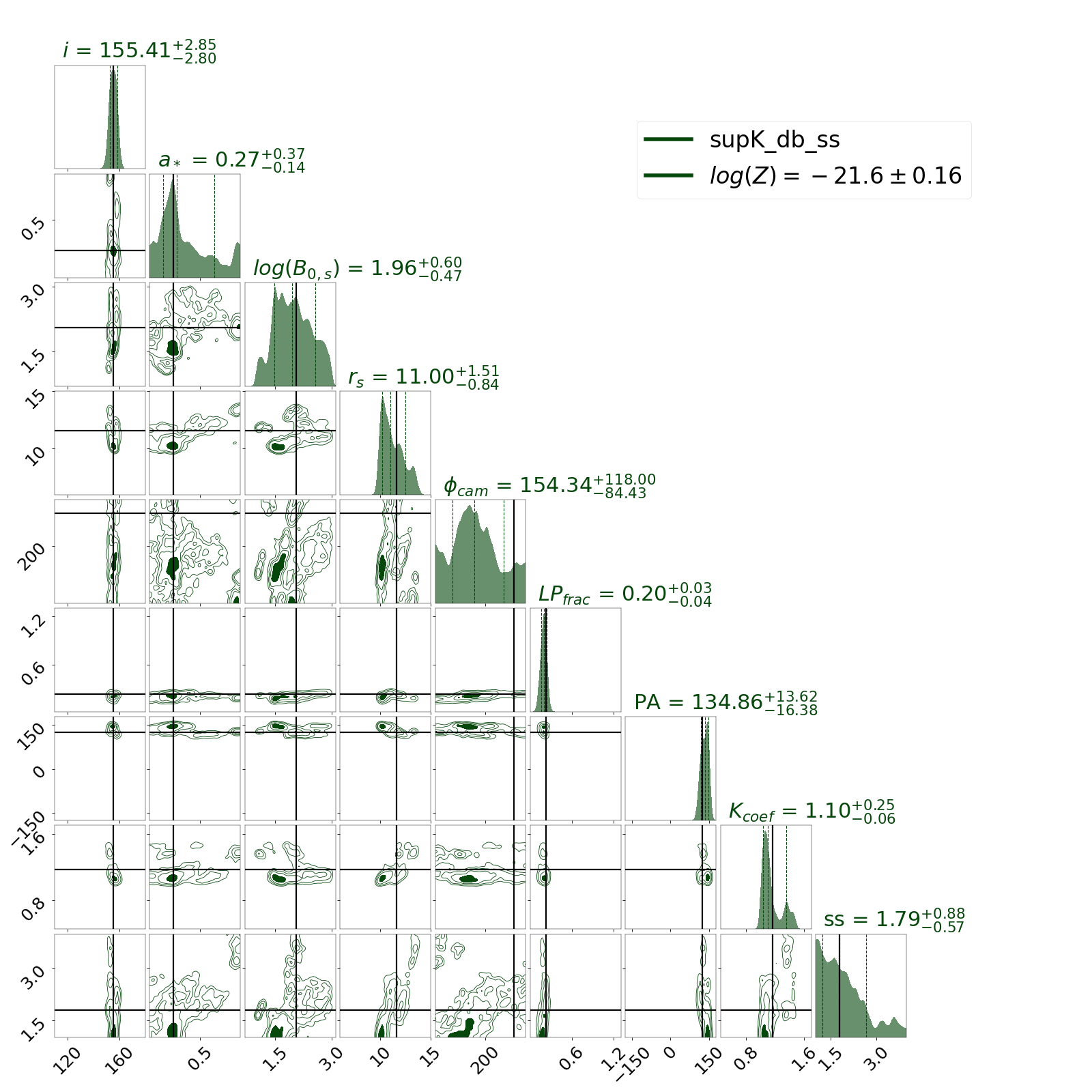} 
    
    \includegraphics[width=0.45\linewidth,trim={0.1cm 0.0cm 0 0.1cm},clip]{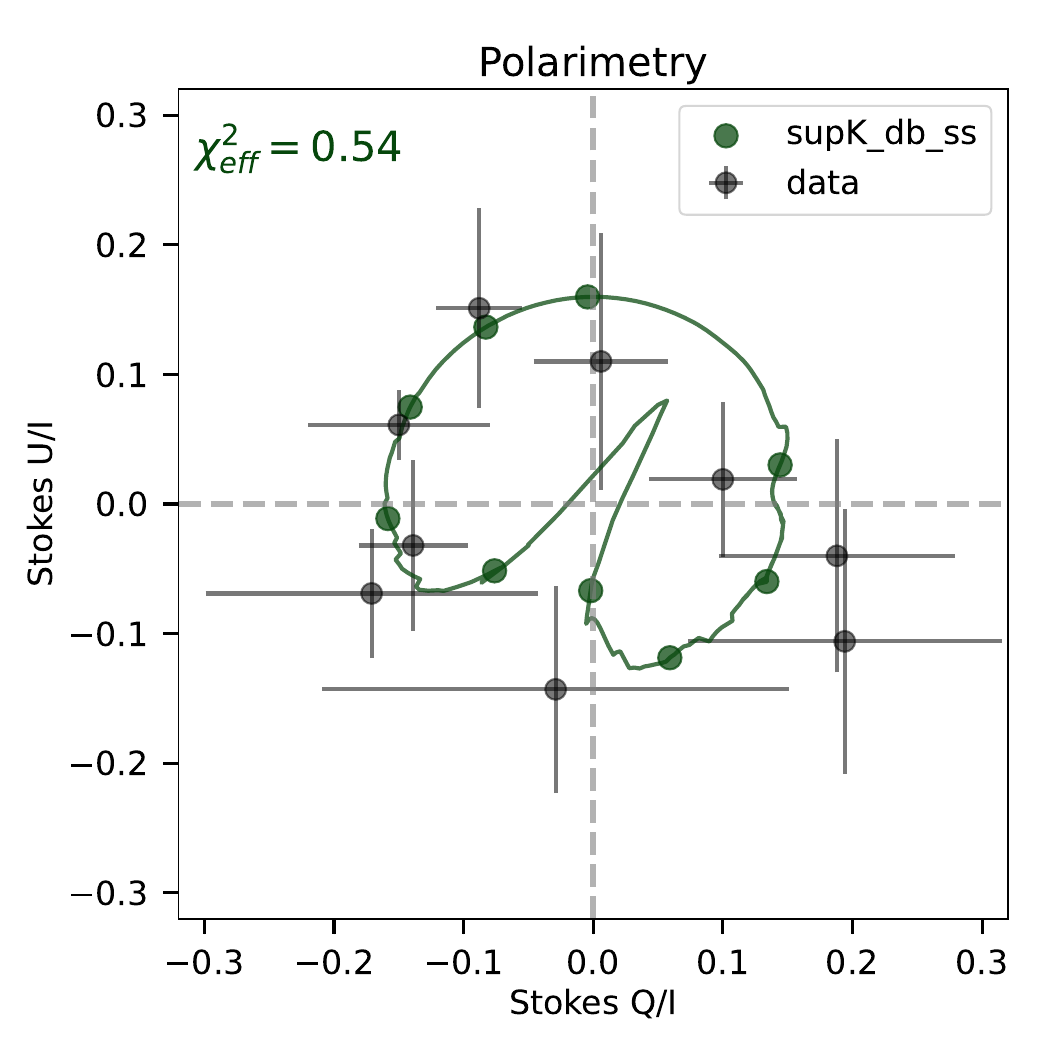} 
    \includegraphics[width=0.461\linewidth,trim={0.1cm 0.0cm 0 0.1cm},clip]{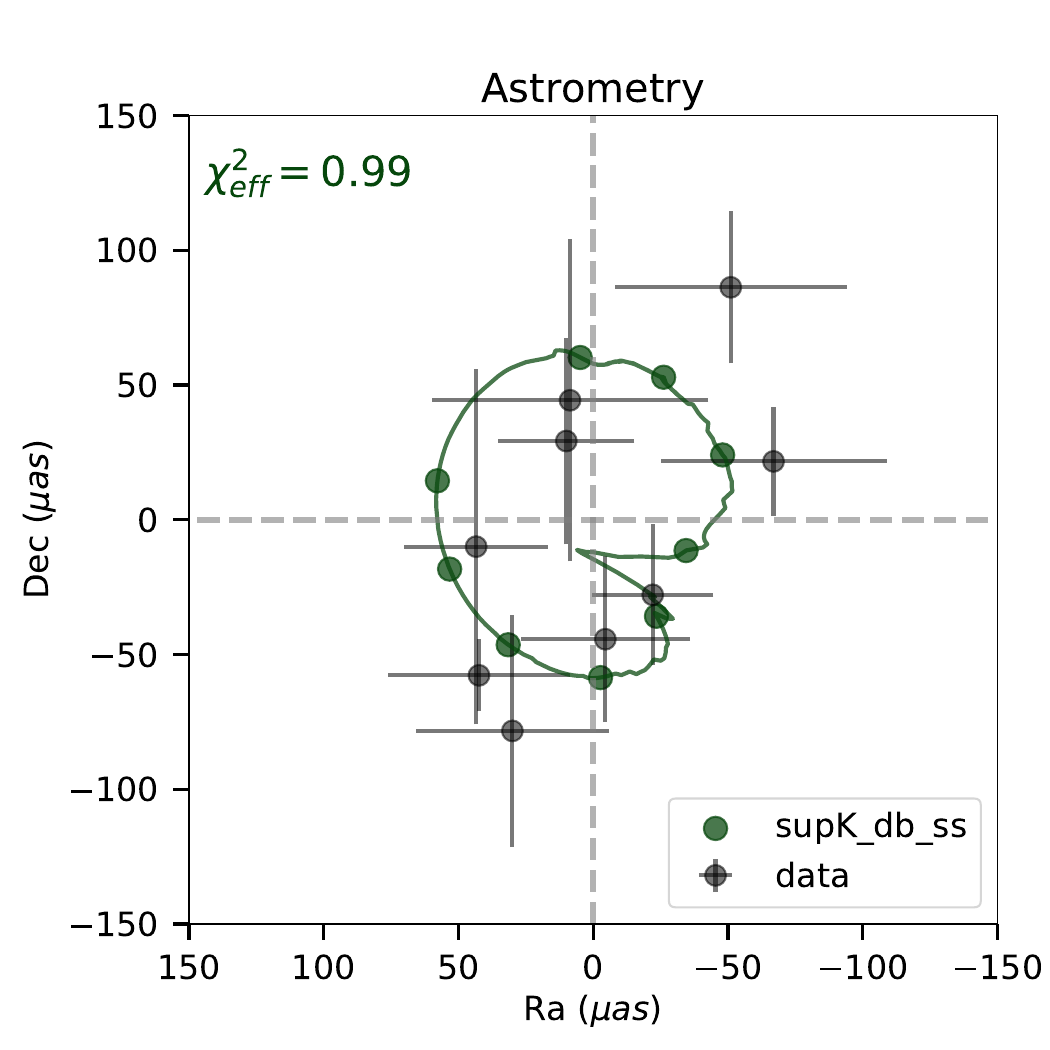} 
    
    \caption{Same as in Fig. \ref{fig:corner_K} but for the model with variable hot spot size supK\_db\_ss.}
    \label{fig:corner_nK_ss}
\end{figure*}

\begin{figure*}[tbh!]
    \centering
    \includegraphics[width=0.781\linewidth,trim={0.2cm 0.0cm 0 0.2cm},clip]{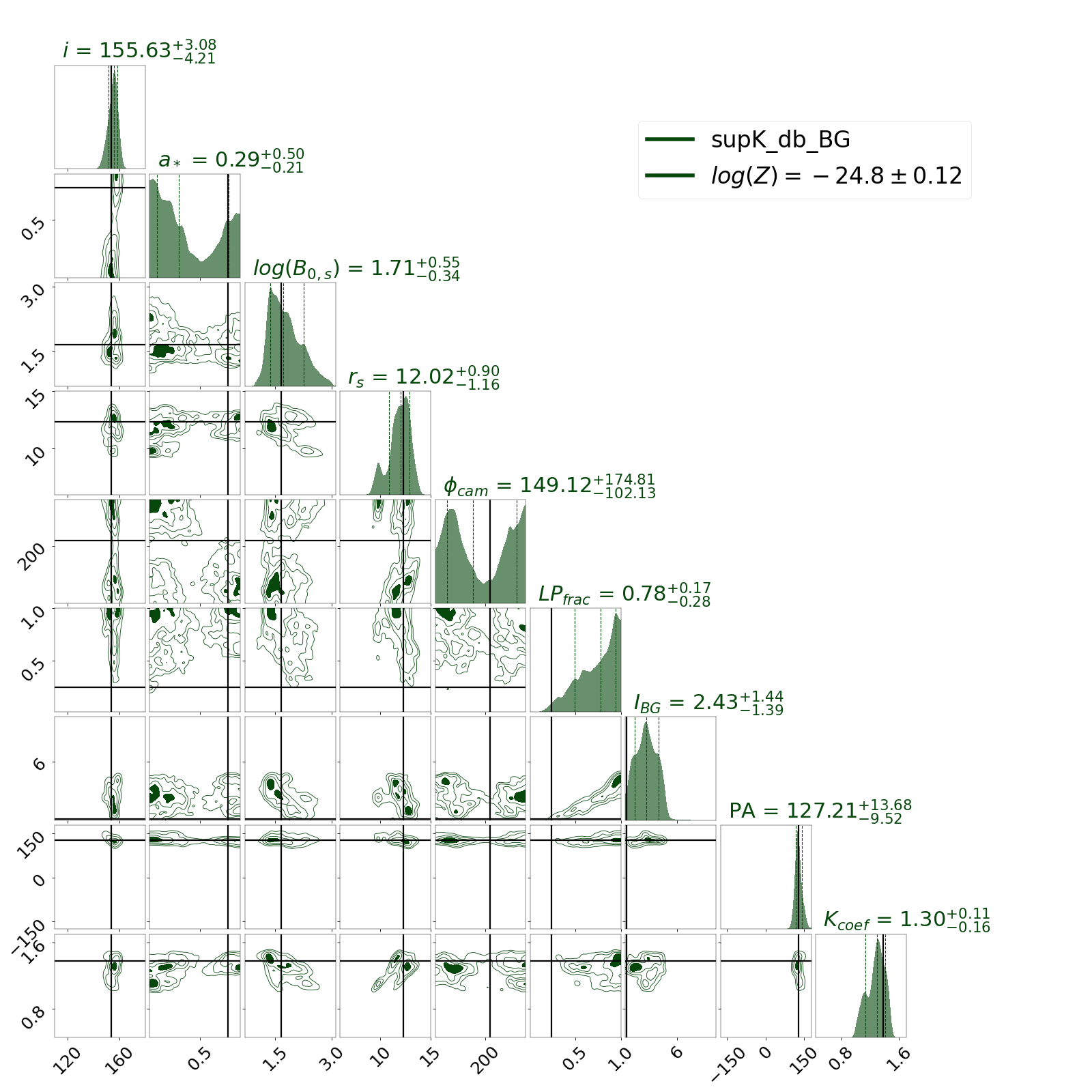} \\

    \includegraphics[width=0.45\linewidth,trim={0.1cm 0.0cm 0 0.1cm},clip]{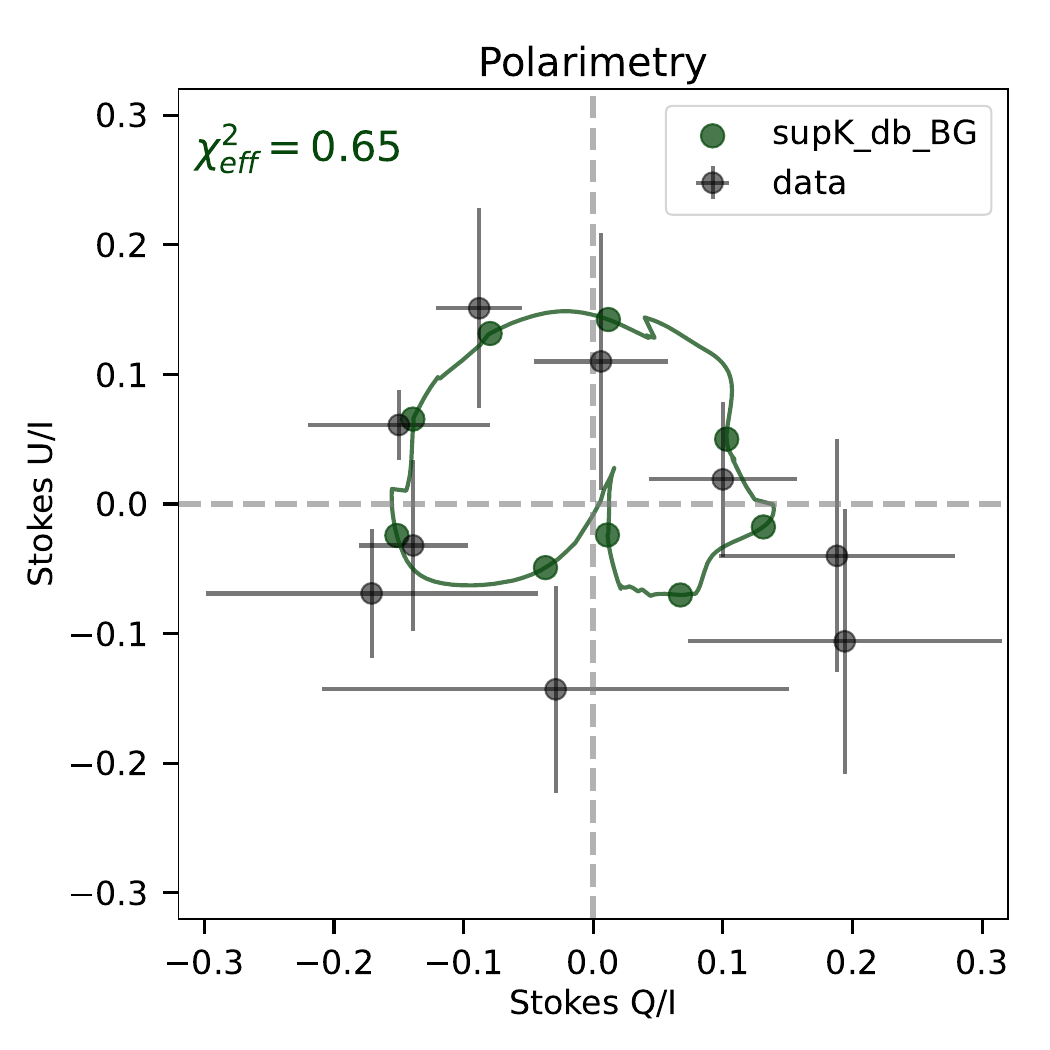} 
    \includegraphics[width=0.461\linewidth,trim={0.1cm 0.0cm 0 0.1cm},clip]{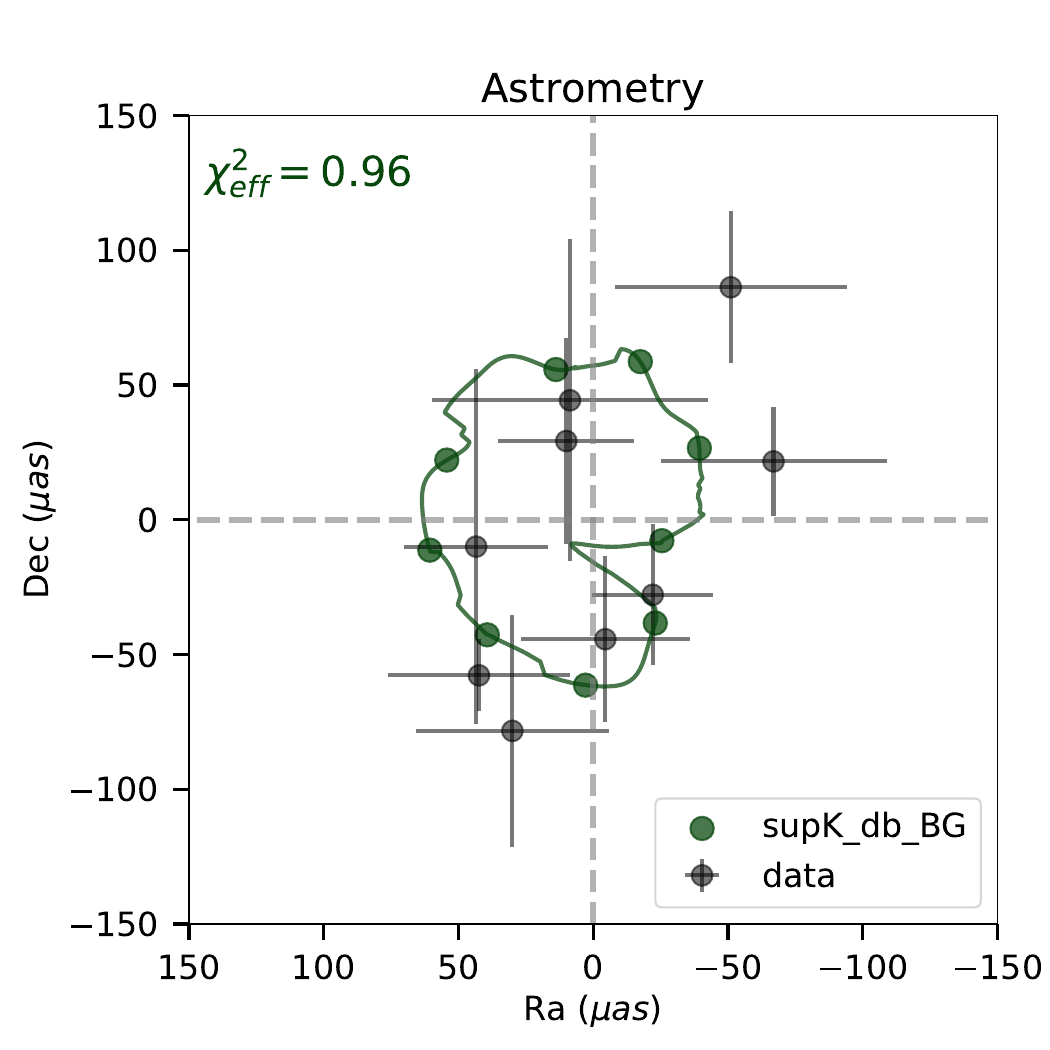} 
    
    \caption{Same as Fig. \ref{fig:corner_K} but for the background component supK\_db\_BG model.}
    \label{fig:corner_bg}
\end{figure*}

\begin{figure*}
    \centering
    \includegraphics[width=0.781\linewidth,trim={0.2cm 0.0cm 0 0.2cm},clip]{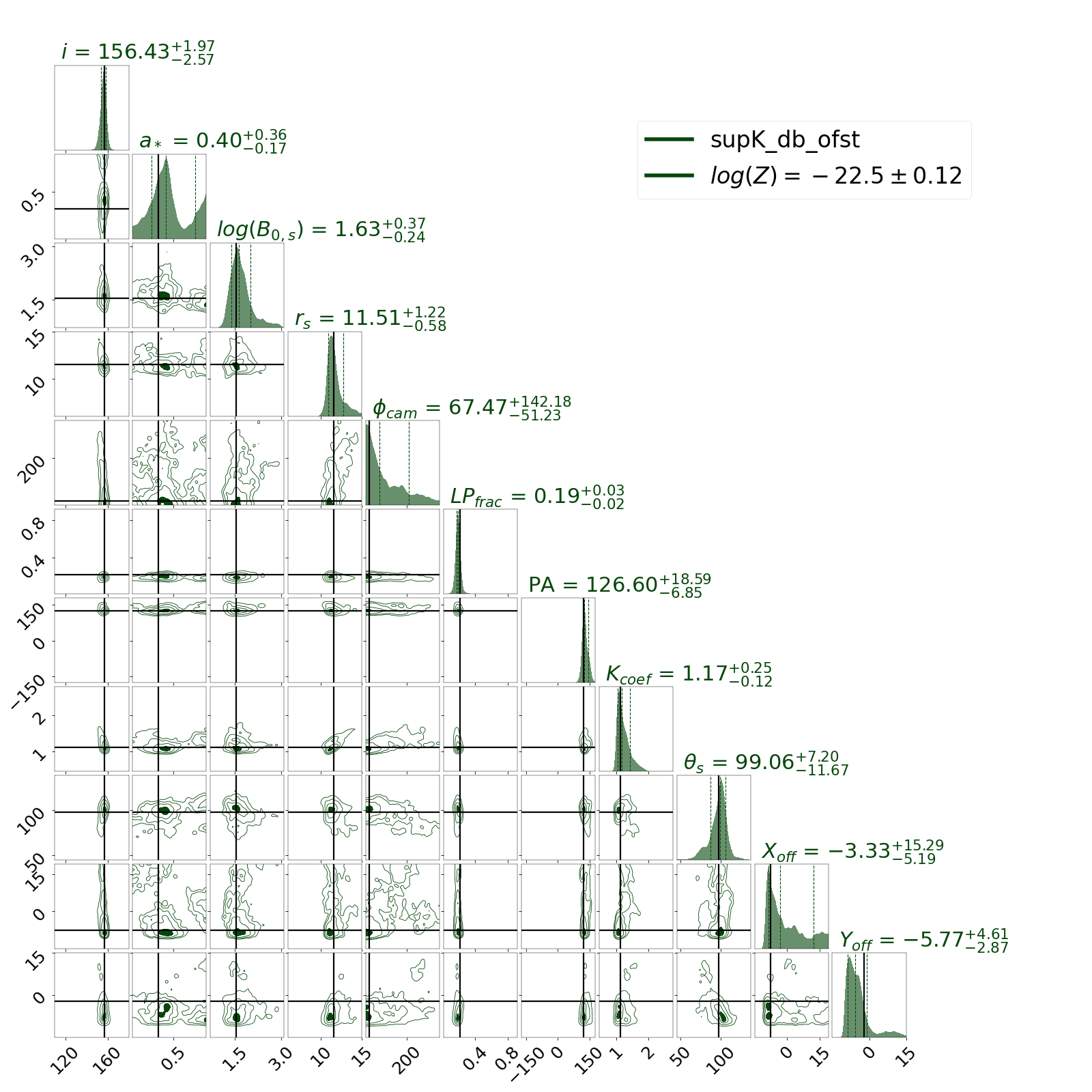} \\

    \includegraphics[width=0.45\linewidth,trim={0.1cm 0.0cm 0 0.1cm},clip]{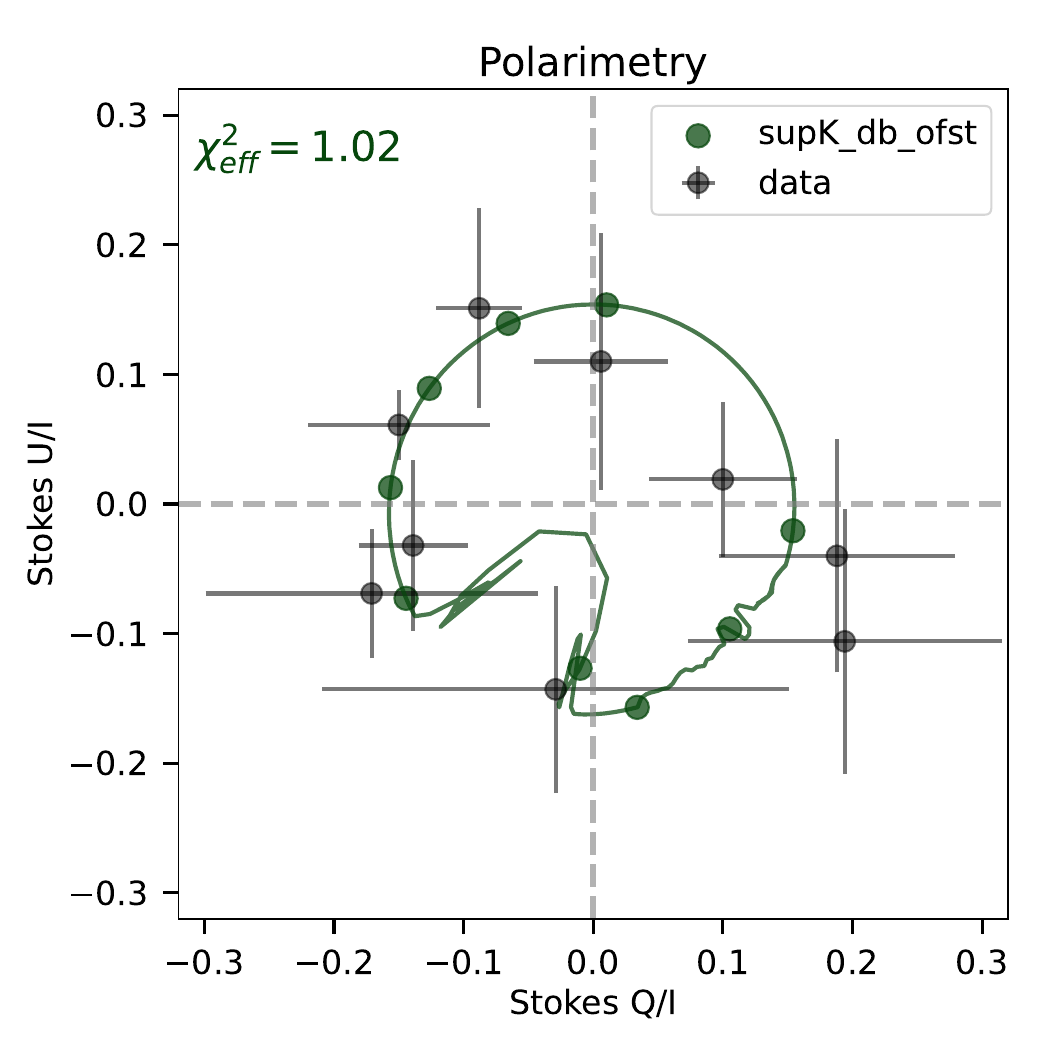} 
    \includegraphics[width=0.461\linewidth,trim={0.1cm 0.0cm 0 0.1cm},clip]{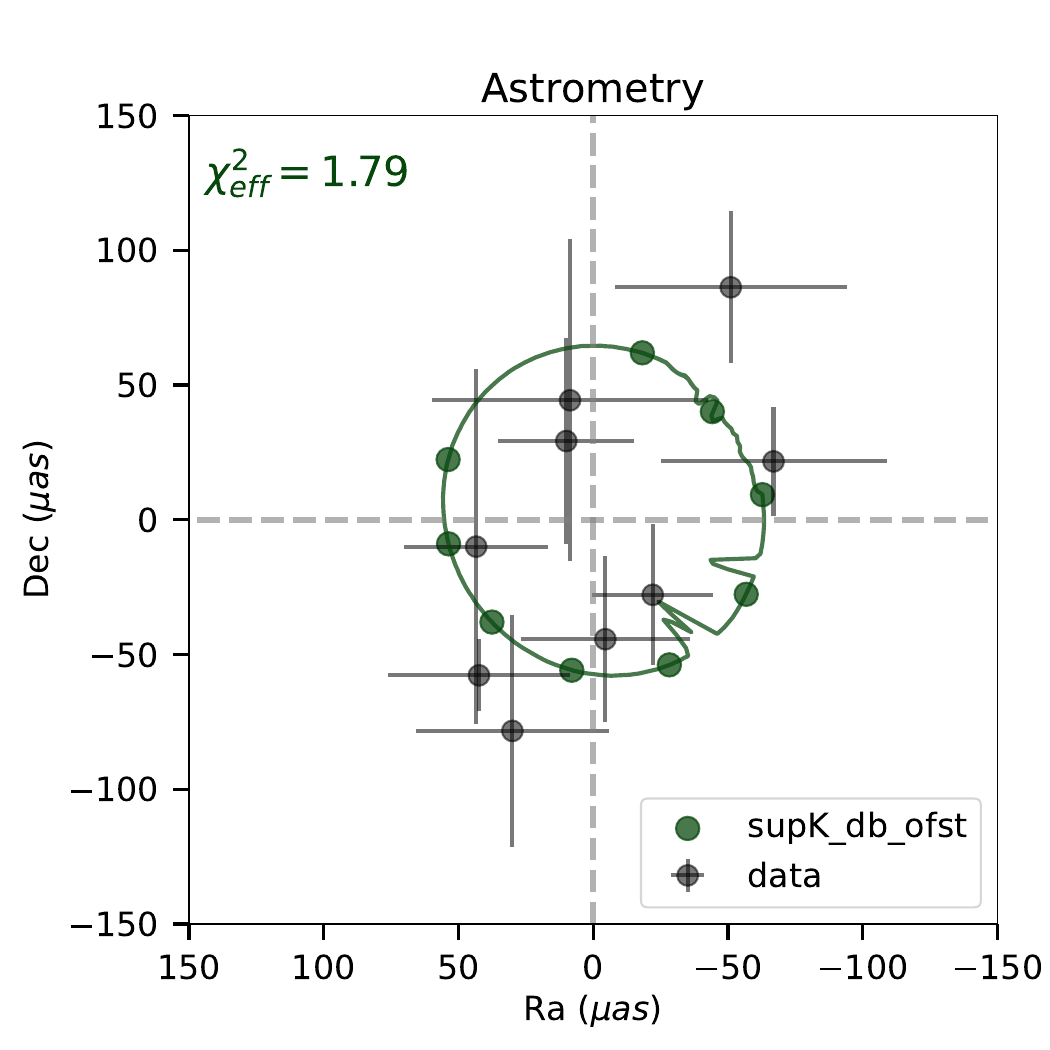} 

     \caption{Same as Fig. \ref{fig:corner_K} but for the off-equatorial orbit model supK\_db\_ofst.}

    \label{fig:corner_th}
\end{figure*}

\end{appendix}

%
%

\end{document}